\documentclass[authoryear,9pt,preprint]{sigplanconf}

\makeatletter
\@ifundefined{lhs2tex.lhs2tex.sty.read}%
  {\@namedef{lhs2tex.lhs2tex.sty.read}{}%
   \newcommand\SkipToFmtEnd{}%
   \newcommand\EndFmtInput{}%
   \long\def\SkipToFmtEnd#1\EndFmtInput{}%
  }\SkipToFmtEnd

\newcommand\ReadOnlyOnce[1]{\@ifundefined{#1}{\@namedef{#1}{}}\SkipToFmtEnd}
\usepackage{amstext}
\usepackage{amssymb}
\usepackage{stmaryrd}
\DeclareFontFamily{OT1}{cmtex}{}
\DeclareFontShape{OT1}{cmtex}{m}{n}
  {<5><6><7><8>cmtex8
   <9>cmtex9
   <10><10.95><12><14.4><17.28><20.74><24.88>cmtex10}{}
\DeclareFontShape{OT1}{cmtex}{m}{it}
  {<-> ssub * cmtt/m/it}{}

\DeclareFontShape{OT1}{cmtt}{bx}{n}
  {<5><6><7><8>cmtt8
   <9>cmbtt9
   <10><10.95><12><14.4><17.28><20.74><24.88>cmbtt10}{}
\DeclareFontShape{OT1}{cmtex}{bx}{n}
  {<-> ssub * cmtt/bx/n}{}

\newcommand{\Conid}[1]{\mathit{#1}}
\newcommand{\Varid}[1]{\mathit{#1}}
\newcommand{\anonymous}{\kern0.06em \vbox{\hrule\@width.5em}}

\newcommand{\bind}{\mathbin{>\!\!\!>\mkern-6.7mu=}}

\usepackage{polytable}

\@ifundefined{mathindent}%
  {\newdimen\mathindent\mathindent\leftmargini}%
  {}%

\def\resethooks{%
  \global\let\SaveRestoreHook\empty
  \global\let\ColumnHook\empty}
\newcommand*{\savecolumns}[1][default]%
  {\g@addto@macro\SaveRestoreHook{\savecolumns[#1]}}
\newcommand*{\restorecolumns}[1][default]%
  {\g@addto@macro\SaveRestoreHook{\restorecolumns[#1]}}
\newcommand*{\aligncolumn}[2]%
  {\g@addto@macro\ColumnHook{\column{#1}{#2}}}

\resethooks

\newcommand{\onelinecommentchars}{\quad-{}- }
\newcommand{\commentbeginchars}{\enskip\{-}
\newcommand{\commentendchars}{-\}\enskip}

\newcommand{\visiblecomments}{%
  \let\onelinecomment=\onelinecommentchars
  \let\commentbegin=\commentbeginchars
  \let\commentend=\commentendchars}

\newcommand{\invisiblecomments}{%
  \let\onelinecomment=\empty
  \let\commentbegin=\empty
  \let\commentend=\empty}

\visiblecomments

\newlength{\blanklineskip}
\newcommand{\hsindent}[1]{\quad}
\let\hspre\empty
\let\hspost\empty

\EndFmtInput
\makeatother
\ReadOnlyOnce{polycode.fmt}%
\makeatletter

\newcommand{\hsnewpar}[1]%
  {{\parskip=0pt\parindent=0pt\par\vskip #1\noindent}}

\newcommand{\hscodestyle}{}

\newcommand{\sethscode}[1]%
  {\expandafter\let\expandafter\hscode\csname #1\endcsname
   \expandafter\let\expandafter\endhscode\csname end#1\endcsname}

  {\par\noindent
   \advance\leftskip\mathindent
   \hscodestyle
   \let\\=\@normalcr
   \let\hspre\(\let\hspost\)%
   \pboxed}%
  {\endpboxed\)%
   \par\noindent
   \ignorespacesafterend}

  {\hsnewpar\abovedisplayskip
   \advance\leftskip\mathindent
   \hscodestyle
   \let\hspre\(\let\hspost\)%
   \pboxed}%
  {\endpboxed%
   \hsnewpar\belowdisplayskip
   \ignorespacesafterend}

  {\hsnewpar\abovedisplayskip
   \advance\leftskip\mathindent
   \hscodestyle
   \let\\=\@normalcr
   \(\pboxed}%
  {\endpboxed\)%
   \hsnewpar\belowdisplayskip
   \ignorespacesafterend}

\newcommand{\plainhs}{\sethscode{plainhscode}}

\plainhs

  {\hsnewpar\abovedisplayskip
   \advance\leftskip\mathindent
   \hscodestyle
   \let\\=\@normalcr
   \(\parray}%
  {\endparray\)%
   \hsnewpar\belowdisplayskip
   \ignorespacesafterend}

  {\parray}{\endparray}

  {\(\parray}{\endparray\)}

\def\codeframewidth{\arrayrulewidth}
\RequirePackage{calc}

  {\parskip=\abovedisplayskip\par\noindent
   \hscodestyle
   \arrayrulewidth=\codeframewidth
   \tabular{@{}|p{\linewidth-2\arraycolsep-2\arrayrulewidth-2pt}|@{}}%
   \hline\framedhslinecorrect\\{-1.5ex}%
   \let\endoflinesave=\\
   \let\\=\@normalcr
   \(\pboxed}%
  {\endpboxed\)%
   \framedhslinecorrect\endoflinesave{.5ex}\hline
   \endtabular
   \parskip=\belowdisplayskip\par\noindent
   \ignorespacesafterend}

\newcommand{\framedhslinecorrect}[2]%
  {#1[#2]}

  {\(\def\column##1##2{}%
   \let\>\undefined\let\<\undefined\let\\\undefined
   \newcommand\>[1][]{}\newcommand\<[1][]{}\newcommand\\[1][]{}%
   \def\fromto##1##2##3{##3}%
   }{\) }%

  {\let\orighscode=\hscode
   \let\origendhscode=\endhscode
   \def\endhscode{\def\hscode{\endgroup\def\@currenvir{hscode}\\}\begingroup}
   \orighscode\def\hscode{\endgroup\def\@currenvir{hscode}}}%
  {\origendhscode
   \global\let\hscode=\orighscode
   \global\let\endhscode=\origendhscode}%

\makeatother
\EndFmtInput
\ReadOnlyOnce{forall.fmt}%
\makeatletter

\let\HaskellResetHook\empty
\newcommand*{\AtHaskellReset}[1]{%
  \g@addto@macro\HaskellResetHook{#1}}
\newcommand*{\HaskellReset}{\HaskellResetHook}

\newcommand\hsforall{\global\let\hsdot=\hsperiodonce}
\newcommand*\hsperiodonce[2]{#2\global\let\hsdot=\hscompose}
\newcommand*\hscompose[2]{#1}

\AtHaskellReset{\global\let\hsdot=\hscompose}

\HaskellReset

\makeatother
\EndFmtInput

\usepackage{amsmath}
\usepackage{amsfonts}
\usepackage{amssymb}
\usepackage{stmaryrd}
\usepackage{proof}
\usepackage{xspace}
\usepackage[
  pdfauthor={Shayan Najd,Sam Lindley,Josef Svenningsson,Philip Wadler},
  pdftitle={Everything old is new again:
            Quoted Domain Specific Languages},
  pagebackref=true,pdftex,backref=none]{hyperref}
\usepackage{tabularx}
\usepackage{graphicx}
\usepackage{url}
\usepackage{color}
\usepackage[usenames,dvipsnames,svgnames,table]{xcolor}
\usepackage{listings}
\newcommand{\todo}[2]{}

\newcommand{\longeq}{=\!=}
\newcommand{\rmdiv}{\mathbin{\textrm{div}}}

\newtheorem{theorem}{Theorem}[section]
\newtheorem{proposition}[theorem]{Proposition}

\newcommand{\flushr}{{}\mbox{~}\hfill}

\newcommand{\arrow}[2]{#1\rightarrow#2}
\newcommand{\product}[2]{#1\times#2}
\newcommand{\coproduct}[2]{#1+#2}

\newcommand{\key}{\mathbf}

\newcommand{\app}{\;}

\newcommand{\expvar}[1]{#1}
\newcommand{\expconst}[2]{#1~#2}

\newcommand{\expabs}[3]{\lambda#1.\,#3}
\newcommand{\expapp}[2]{#1 \app #2}
\newcommand{\explet}[3]{\key{let}\ #1=#2\ \key{in}\ #3}
\newcommand{\exppair}[2]{(#1, #2)}
\newcommand{\expfst}[1]{\key{fst}~#1}
\newcommand{\expsnd}[1]{\key{snd}~#1}
\newcommand{\expinl}[2]{\key{inl}~#1}
\newcommand{\expinr}[2]{\key{inr}~#2}
\newcommand{\expcase}[5]
  {\key{case}\ #1\ \key{of}\ \{\key{inl}\ #2.\,#3;\; \key{inr}\ #4.\,#5\}}

\newcommand{\subst}[3]{#1[#2:=#3]}
\newcommand{\fv}[1]{\mathit{FV}(#1)}
\newcommand{\rewrite}[1]{\mathbin{\mapsto_{#1}}}
\newcommand{\hole}{[~]}

\newcommand{\figterm}{
\begin{figure*}[t]
\[
\begin{array}{l@{\quad}rcl}
\text{Types} & A,B,C & ::=&
  \iota           \mid
  \arrow{A}{B}    \mid
  \product{A}{B}  \mid
  \coproduct{A}{B}
\\[1ex]
\text{Terms} & L,M,N & ::= &
  \expvar{x}              	\mid
  \expconst{c}{\overline{M}}	\mid
  \expabs{x}{A}{N}              \mid
  \expapp{L}{M}                 \mid
  \explet{x}{M}{N}              \mid
  \exppair{M}{N}                \mid
  \expfst{L}                   	\mid
  \expsnd{L}                    \mid \\&&&
  \expinl{M}{\tm{B}}		\mid
  \expinr{\tm{A}}{N}           	\mid
  \expcase{L}{x}{M}{y}{N}
\\[1ex]
\text{Values} & V,W & ::= &
  \expvar{x}       \mid
  \expabs{x}{A}{N} \mid
  \exppair{V}{W}   \mid
  \expinl{V}{B}    \mid
  \expinr{A}{W}
\end{array}
\]

\caption{Types and Terms}
\label{fig:term}
\end{figure*}
}

\newcommand{\fignorm}{
\begin{figure*}[t]
Phase 1 (let-insertion)
\[
  \begin{array}{lcl}
   F &\mathbin{::=}&
       \expapp{\hole}{M}                                \mid
       \expapp{V}{\hole}                                \mid
       \exppair{\hole}{M}                               \mid
       \exppair{V}{\hole}                               \mid
       \expfst{\hole}                                   \mid
       \expsnd{\hole}                                   \mid 
       \expinl{\hole}{B}                                \mid
       \expinr{A}{\hole}                                \mid
       \expcase{\hole}{x}{M}{y}{N} \\
 \end{array}
\]%
\[
\begin{array}{lllll}
(\mathit{let})
& F[M]
& \rewrite{1}
& \explet{x}{M}{F[x]},
& \text{$x$ fresh, $M$ not a value}
\end{array}
\]

\vspace{2ex}

Phase 2 (symbolic evaluation)
\[
G \mathbin{::=}
    \explet{x}{\hole}{N}
\]
\[
\begin{array}{lllll}

(\kappa.{\mathit{let}})
& G[\explet{x}{M}{N}]
& \rewrite{2}
& \explet{x}{M}{G[N]},
& x \notin \fv{G} \\

(\kappa.{\mathit{case}})
& G[\expcase{V}{x}{M}{y}{N}]
& \rewrite{2}
& \expcase{V}{x}{G[M]}{y}{G[N]},
& x,y \notin \fv{G} \\

(\beta.{\rightarrow})
& \expapp{(\expabs{x}{A}{N})}{V}
& \rewrite{2}
& \subst{N}{x}{V} \\

(\beta.{\times_1})
& \expfst{\exppair{V}{W}}
& \rewrite{2}
& V \\

(\beta.{\times_2})
& \expsnd{\exppair{V}{W}}
& \rewrite{2}
& W \\

(\beta.{+_1})
& \expcase{(\expinl{V}{B})}{x}{M}{y}{N}
& \rewrite{2}
& \subst{M}{x}{V} \\

(\beta.{+_2})
& \expcase{(\expinr{A}{W})}{x}{M}{y}{N}
& \rewrite{2}
& \subst{N}{y}{W} \\

(\beta.{\mathit{let}})
& \explet{x}{V}{N}
& \rewrite{2}
& \subst{N}{x}{V} \\
\end{array}
\]

\vspace{2ex}

Phase 3 (garbage collection)
\[
\begin{array}[t]{@{}llll@{\quad}l@{}}

(\mathit{need})
& \explet{x}{M}{N}
& \rewrite{3}
& N,
& x \notin \fv{N}\\
\end{array}
\]

\caption{Normalisation Rules}
\label{fig:norm}
\end{figure*}
}

\conferenceinfo{ICFP 2015}
               {August 31--September 2, 2015, Vancouver, Canada.}
\copyrightyear{2015}
\copyrightdata{...}
\doi{...}

\begin{document}


\title{Everything old is new again:\\
       Quoted Domain Specific Languages}

\authorinfo{Shayan Najd \and Sam Lindley}
           {The University of Edinburgh}
           {\{sh.najd,sam.lindley\}@ed.ac.uk}
\authorinfo{Josef Svenningsson}
           {Chalmers University of Technology}
           {josefs@chalmers.se}
\authorinfo{Philip Wadler}
           {The University of Edinburgh}
           {wadler@inf.ed.ac.uk}

\maketitle

\begin{abstract}

We describe a new approach to domain specific languages (DSLs),
called Quoted DSLs (QDSLs), that resurrects two old ideas:
quotation, from McCarthy's Lisp of 1960, and the subformula
property, from Gentzen's natural deduction of 1935.  Quoted terms
allow the DSL to share the syntax and type system of the host
language.  Normalising quoted terms ensures the subformula property,
which guarantees that one can use higher-order types in the source
while guaranteeing first-order types in the target, and enables
using types to guide fusion.  We test our ideas by re-implementing
Feldspar, which was originally implemented as an Embedded DSL
(EDSL), as a QDSL; and we compare the QDSL and EDSL variants.

\end{abstract}




\section{Introduction}
\label{sec:introduction}



\begin{quote}
Don't throw the past away \\
You might need it some rainy day \\
Dreams can come true again \\
When everything old is new again \\
\flushr -- Peter Allen and Carole Sager
\end{quote}

Implementing domain-specific languages (DSLs) via quotation is one of
the oldest ideas in computing, going back at least to McCarthy's Lisp,
which was introduced in 1960 and had macros as early as 1963.  Today,
a more fashionable technique is Embdedded DSLs (EDSLs), which may use
shallow embedding, deep embedding, or a combination of the two. Our
goal in this paper is to reinvigorate the idea of building DSLs via
quotation, by introducing an approach we dub Quoted DSLs
(QDSLs).  A key feature of QDSLs is the use of normalisation to ensure
the subformula property, first proposed by Gentzen in 1935.

\vspace{2ex}
\begin{quote}
Imitation is the sincerest of flattery. \\
\flushr --- Charles Caleb Colton
\end{quote}

\citet{cheney:linq} describe a DSL for language-integrated query in
F\# that translates into SQL.  Their technique depends on quotation,
normalisation of quoted terms, and the subformula property---an
approach which we here dub QDSL.  They conjecture that other DSLs might
benefit from the same technique, particularly those that perform
staged computation, where host code at generation-time computes target
code to be executed at run-time.

Generality starts at two.  Here we test the conjecture of
\citet{cheney:linq} by reimplementing the EDSL Feldspar
\citep{FELDSPAR} as a QDSL.  We describe the key features of the
design, and show that the performance of the two versions is
comparable.  We compare the QDSL and EDSL variants of Feldspar, and
assess the tradeoffs between the two approaches.

\citet{Davies-Pfenning-2001} also suggest
quotation as a foundation for staged computation, and note a
propositions-as-types connection between quotation and a modal logic;
our type \ensuremath{\Conid{Qt}\;\Varid{a}} corresponds to their type $\bigcirc a$.  They also
mention in passing the utility of normalising quoted terms, although
they do not mention the subformula property.

The .NET Language-Integrated Query (LINQ) framework as used in C\# and
F\# \citep{csharplinq,fsharplinq}, and the Lightweight Modular Staging
(LMS) framework as used in Scala \citep{scalalms}, exhibit overlap
with the techniques described here.  Notably, they use quotation to
represent staged DSL programs, and they make use to a greater or
lesser extent of normalisation.  In F\# LINQ quotation is indicated in
the normal way (by writing quoted programs inside special symbols),
while in C\# LINQ and Scala LMS quotation is indicated by type
inference (quoted terms are given a special type).


\vspace{2ex}
\begin{quote}
Perhaps we may express the essential properties of such a normal proof
by saying: it is not roundabout. \\
\flushr --- Gerhard Gentzen
\end{quote}

Our approach exploits the fact that normalised terms satisfy the
subformula property, first introduced in the context
of natural deduction by \citet{Gentzen-1935},
and improved by \citet{Prawitz-1965}.

The subformulas of a formula are its subparts;
for instance, the subformulas of \ensuremath{\Conid{A}\to \Conid{B}} are the formula
itself and the subformulas of \ensuremath{\Conid{A}} and \ensuremath{\Conid{B}}.
The subformula property states that every proof can be put into a
normal form where the only propositions that appear in the proof are
subformulas of the hypotheses and conclusion of the proof. Applying
the principle of Propositions as Types
\citep{Howard-1980,Wadler-2015}, the subformula property states that
every lambda term can be put into a normal form where the only types
that appear in the term are subformulas of the types of the free
variables and the type of the term itself.

The subformula property provides users of the
DSL with useful guarantees, such as the following:
\begin{itemize}

\item they may write higher-order terms
while guaranteeing to generate first-order code;

\item they may write a sequence of loops over arrays
while guaranteeing to generate code that fuses those loops;

\item they may write intermediate terms with nested collections
while guaranteeing to generate code that operates on flat data.

\end{itemize}
The first and second are used in this paper, and are key
to generating C; while the
first and third are used by \citet{cheney:linq}
and are key to generating SQL.

The subformula property is closely related to conservativity.  A
conservativity result expresses that adding a feature to a system of
logic, or to a programming language, does not make it more expressive.
Consider intuitionistic logic with conjunction; conservativity states
that adding implication to this logic proves no additional theorems
that can be stated in the original logic.
Such a conservativity result
is an immediate consequence of the subformula property; since the
hypotheses and conjuction of the proof only mention conjunction, any
proof, even if it uses implication, can be put into a normal form that
only uses conjunction. Equivalently, any lambda calculus term that
mentions only pair types in its free variables and result, even if it
uses functions, can be put in a normal form that only uses pairs. Such
a result is related to the first bullet point above; see
Proposition~\ref{prop:fo} in Section~\ref{sec:subformula}.

As another example, the third bullet point above corresponds to a
standard conservativity result for databases, namely that nested
queries are no more expressive than flat queries \citep{Wong-1996}.
This conservativity result, as implied by the subformula property, is
used by \citet{cheney:linq} to show that queries that use intermediate
nesting can be translated to SQL, which only queries flat tables and
does not support nesting of data.

The subformula property holds only for terms in normal form.  Previous
work, such as \citep{cheney:linq} uses call-by-name normalisation
algorithm that may cause
computations to be repeated.  Here we present call-by-value and
call-by-need normalisation algorithms that guarantee to preserve
sharing of computations. We also present a sharpened version of
the subformula property, which we apply to characterise
the circumstances under which a QDSL may guarantee
to generate first-order code.

\vspace{2ex}
\begin{quote}
Good artists copy, great artists steal. \flushr --- Picasso
\end{quote}

EDSL is great in part because it steals the type system of its host
language. Arguably, QDSL is greater because it steals the type system,
the syntax, and the normalisation rules of its host language.

In theory, an EDSL should also steal the syntax of its host language,
but in practice the theft is often only partial.  For instance, an
EDSL such as Feldspar \citep{FELDSPAR} or Nikola \citep{NIKOLA},
when embedded in Haskell, can exploit
overloading so that arithmetic operations in both
languages appear identical, but the same is not true of comparison or
conditionals.
In QDSL, the syntax of the host and embedded languages
is identical. For instance, this paper presents a QDSL variant of
Feldspar, again in Haskell, where arithmetic, comparison, and
conditionals are all represented by quoted terms, and hence identical
to the host.

An EDSL may also steal the normalisation rules of its host language,
using evaluation in the host to normalise terms of the
target, but again the theft is often only partial. Section~\ref{sec:qdsl-vs-edsl}
compares QDSL and EDSL variants of Feldspar. In the first example, it
is indeed the case that the EDSL achieves by evaluation of host terms
what the QDSL achieves by normalisation of quoted terms.  However, in
other cases, the EDSL must perform normalisation of the deep embedding
corresponding to how the QDSL normalises quoted terms.

\vspace{2ex}
\begin{quote}
Try to give all of the information to help others to judge the value
of your contribution; not just the information that leads to judgment
in one particular direction or another. \\
\flushr --- Richard Feynman
\end{quote}

The subformula property depends on normalisation, but normalisation
may lead to exponential blowup in the size of the normalised
code when there are nested conditionals;
and hyperexponential blowup in recondite cases involving higher-order functions.
We explain how uninterpreted constants allow the user to
control where normalisation does and does not occur, while still
maintaining the subformula property. Future work is required to
consider trade-offs between full normalisation as required for
the subformula property and special-purpose normalisation as used
in many DSLs; possibly a combination of both will prove fruitful.

Some researchers contend an essential property of a DSL
which generates target code is that every type-correct term
should successfully generate code in the target language. EDSL Feldspar
satisfies this property; but neither
P-LINQ of \citet{cheney:linq} nor QDSL Feldspar
satisfy this property, since the user is required to eyeball quoted
code to ensure it mentions only permitted operators. If this is
thought too onerous, it is possible to ensure the property with
additional preprocessing.

\vspace{2ex}
\begin{quote}
This is the short and the long of it. \flushr --- Shakespeare
\end{quote}

The contributions of this paper are:
\begin{itemize}

\item To introduce QDSLs as an approach to building DSLs based on
  quotation, normalisation of quoted terms, and the subformula property
  by presenting the design of a QDSL variant of Feldspar
  (Section~\ref{sec:qfeldspar}).

\item To measure QDSL and EDSL implementations of Feldspar,
  and show they offer comparable performance
  (Section~\ref{sec:implementation}).

\item To present normalisation algorithms for
  call-by-value and call-by-need that preserve sharing, and to
  formulate a sharpened version of the subformula property and apply
  it to characterise when higher-order terms normalise to first-order
  form (Section~\ref{sec:subformula}).

\item To compare the QDSL variant of Feldspar with the deep and
  shallow embedding approach used in the EDSL variant of Feldspar,
  and show they offer tradeoffs with regard to ease of use
  (Section~\ref{sec:qdsl-vs-edsl}).

\end{itemize}
Section~\ref{sec:related} describes related work, and
Section~\ref{sec:conclusion} concludes.

Our QDSL and EDSL variants of Feldspar and benchmarks are
available at \url{https://github.com/shayan-najd/QFeldspar}.

\section{Feldspar as a QDSL}
\label{sec:qfeldspar}

Feldspar is an EDSL for writing signal-processing software, that
generates code in C \citep{FELDSPAR}. We present a variant,
QDSL~Feldspar, that follows the structure of the previous design closely,
but using the methods of QDSL rather than EDSL.
Section~\ref{sec:qdsl-vs-edsl} compares the QDSL and EDSL designs.


\subsection{The top level}
\label{subsec:top}

In QDSL~Feldspar, our goal is to translate a quoted term to C code.
The top-level function has the type:
\begin{hscode}\SaveRestoreHook
\column{B}{@{}>{\hspre}l<{\hspost}@{}}%
\column{E}{@{}>{\hspre}l<{\hspost}@{}}%
\>[B]{}\Varid{qdsl}\mathbin{::}(\Conid{Rep}\;\Varid{a},\Conid{Rep}\;\Varid{b})\Rightarrow \Conid{Qt}\;(\Varid{a}\to \Varid{b})\to \Conid{C}{}\<[E]%
\ColumnHook
\end{hscode}\resethooks
Here \ensuremath{\Conid{Qt}\;\Varid{a}} represents a Haskell term of type \ensuremath{\Varid{a}}, its
\emph{quoted} representation, and type \ensuremath{\Conid{C}} represents code in C.
The top-level function expects a quoted term representing
a function from type \ensuremath{\Varid{a}} to type \ensuremath{\Varid{b}}, and returns C code
that computes the function.

Not all types representable in Haskell are easily representable in C.
For instance, we do not wish our target C code to manipulate higher-order functions.
The argument
type \ensuremath{\Varid{a}} and result type \ensuremath{\Varid{b}} of the main function must be representable,
which is indicated by the type-class restrictions \ensuremath{\Conid{Rep}\;\Varid{a}} and \ensuremath{\Conid{Rep}\;\Varid{b}}.
Representable types include integers, floats, and pairs where the components
are both representable.
\begin{hscode}\SaveRestoreHook
\column{B}{@{}>{\hspre}l<{\hspost}@{}}%
\column{E}{@{}>{\hspre}l<{\hspost}@{}}%
\>[B]{}\mathbf{instance}\;\Conid{Rep}\;\Conid{Int}{}\<[E]%
\\
\>[B]{}\mathbf{instance}\;\Conid{Rep}\;\Conid{Float}{}\<[E]%
\\
\>[B]{}\mathbf{instance}\;(\Conid{Rep}\;\Varid{a},\Conid{Rep}\;\Varid{b})\Rightarrow \Conid{Rep}\;(\Varid{a},\Varid{b}){}\<[E]%
\ColumnHook
\end{hscode}\resethooks

\subsection{A first example}
\label{subsec:power}

Let's begin with the ``hello world'' of program generation,
the power function.
Since division by zero is undefined, we
arbitrarily choose that raising zero to a negative power yields zero.
Here is an optimised power function represented using QDSL:
\begin{hscode}\SaveRestoreHook
\column{B}{@{}>{\hspre}l<{\hspost}@{}}%
\column{3}{@{}>{\hspre}l<{\hspost}@{}}%
\column{5}{@{}>{\hspre}l<{\hspost}@{}}%
\column{26}{@{}>{\hspre}l<{\hspost}@{}}%
\column{32}{@{}>{\hspre}l<{\hspost}@{}}%
\column{E}{@{}>{\hspre}l<{\hspost}@{}}%
\>[B]{}\Varid{power}\mathbin{::}\Conid{Int}\to \Conid{Qt}\;(\Conid{Float}\to \Conid{Float}){}\<[E]%
\\
\>[B]{}\Varid{power}\;\Varid{n}\mathrel{=}{}\<[E]%
\\
\>[B]{}\hsindent{3}{}\<[3]%
\>[3]{}\mathbf{if}\;\Varid{n}\mathbin{<}\mathrm{0}\;\mathbf{then}{}\<[E]%
\\
\>[3]{}\hsindent{2}{}\<[5]%
\>[5]{}[\mskip1.5mu ||\lambda \Varid{x}\to \mathbf{if}\;\Varid{x}\longeq \mathrm{0}\;{}\<[26]%
\>[26]{}\mathbf{then}\;{}\<[32]%
\>[32]{}\mathrm{0}{}\<[E]%
\\
\>[26]{}\mathbf{else}\;{}\<[32]%
\>[32]{}\mathrm{1}\mathbin{/}(\mathbin{\$\$}(\Varid{power}\;(\mathbin{-}\Varid{n}))\;\Varid{x})||\mskip1.5mu]{}\<[E]%
\\
\>[B]{}\hsindent{3}{}\<[3]%
\>[3]{}\mathbf{else}\;\mathbf{if}\;\Varid{n}\longeq \mathrm{0}\;\mathbf{then}{}\<[E]%
\\
\>[3]{}\hsindent{2}{}\<[5]%
\>[5]{}[\mskip1.5mu ||\lambda \Varid{x}\to \mathrm{1}||\mskip1.5mu]{}\<[E]%
\\
\>[B]{}\hsindent{3}{}\<[3]%
\>[3]{}\mathbf{else}\;\mathbf{if}\;\Varid{even}\;\Varid{n}\;\mathbf{then}{}\<[E]%
\\
\>[3]{}\hsindent{2}{}\<[5]%
\>[5]{}[\mskip1.5mu ||\lambda \Varid{x}\to \mathbf{let}\;\Varid{y}\mathrel{=}\mathbin{\$\$}(\Varid{power}\;(\Varid{n}\rmdiv\mathrm{2}))\;\Varid{x}\;\mathbf{in}\;\Varid{y}\times\Varid{y}||\mskip1.5mu]{}\<[E]%
\\
\>[B]{}\hsindent{3}{}\<[3]%
\>[3]{}\mathbf{else}{}\<[E]%
\\
\>[3]{}\hsindent{2}{}\<[5]%
\>[5]{}[\mskip1.5mu ||\lambda \Varid{x}\to \Varid{x}\times(\mathbin{\$\$}(\Varid{power}\;(\Varid{n}\mathbin{-}\mathrm{1}))\;\Varid{x})||\mskip1.5mu]{}\<[E]%
\ColumnHook
\end{hscode}\resethooks
The typed quasi-quoting mechanism of Template Haskell is used to
indicate which code executes at which time.  Unquoted code executes at
generation-time while quoted code executes at run-time. Quoting is
indicated by \ensuremath{[\mskip1.5mu \mathbin{||...||}\mskip1.5mu]} and unquoting by \ensuremath{\mathbin{\$\$}(\cdots)}.

Invoking \ensuremath{\Varid{qdsl}\;(\Varid{power}\;(\mathbin{-}\mathrm{6}))} generates code to raise a number to its \ensuremath{\mathbin{-}\mathrm{6}} power.
Evaluating \ensuremath{\Varid{power}\;(\mathbin{-}\mathrm{6})} yields the following:
\begin{hscode}\SaveRestoreHook
\column{B}{@{}>{\hspre}l<{\hspost}@{}}%
\column{33}{@{}>{\hspre}l<{\hspost}@{}}%
\column{35}{@{}>{\hspre}l<{\hspost}@{}}%
\column{37}{@{}>{\hspre}l<{\hspost}@{}}%
\column{39}{@{}>{\hspre}l<{\hspost}@{}}%
\column{E}{@{}>{\hspre}l<{\hspost}@{}}%
\>[B]{}[\mskip1.5mu ||{}\<[33]%
\>[33]{}\lambda \Varid{x}\to \mathbf{if}\;\Varid{x}\longeq \mathrm{0}\;\mathbf{then}\;\mathrm{0}\;\mathbf{else}\;\mathrm{1}\mathbin{/}{}\<[E]%
\\
\>[33]{}\hsindent{2}{}\<[35]%
\>[35]{}(\lambda \Varid{x}\to \mathbf{let}\;\{\mskip1.5mu \Varid{y}\mathrel{=}(\lambda \Varid{x}\to \Varid{x}\times{}\<[E]%
\\
\>[35]{}\hsindent{2}{}\<[37]%
\>[37]{}(\lambda \Varid{x}\to \mathbf{let}\;\{\mskip1.5mu \Varid{y}\mathrel{=}(\lambda \Varid{x}\to \Varid{x}\times(\lambda \Varid{x}\to \mathrm{1})\;\Varid{x})\;\Varid{x}\mskip1.5mu\}{}\<[E]%
\\
\>[37]{}\hsindent{2}{}\<[39]%
\>[39]{}\mathbf{in}\;\Varid{y}\times\Varid{y})\;\Varid{x})\;\Varid{x}\mskip1.5mu\}\;\mathbf{in}\;\Varid{y}\times\Varid{y})\;\Varid{x}||\mskip1.5mu]{}\<[E]%
\ColumnHook
\end{hscode}\resethooks
Normalising as described in Section~\ref{sec:subformula},
with variables renamed for readability, yields the following:
\begin{hscode}\SaveRestoreHook
\column{B}{@{}>{\hspre}l<{\hspost}@{}}%
\column{14}{@{}>{\hspre}l<{\hspost}@{}}%
\column{27}{@{}>{\hspre}c<{\hspost}@{}}%
\column{27E}{@{}l@{}}%
\column{42}{@{}>{\hspre}l<{\hspost}@{}}%
\column{E}{@{}>{\hspre}l<{\hspost}@{}}%
\>[B]{}[\mskip1.5mu ||\lambda \Varid{u}\to {}\<[42]%
\>[42]{}\mathbf{if}\;\Varid{u}\longeq \mathrm{0}\;\mathbf{then}\;\mathrm{0}\;\mathbf{else}{}\<[E]%
\\
\>[B]{}\hsindent{14}{}\<[14]%
\>[14]{}\mathbf{let}\;\Varid{v}\mathrel{=}\Varid{u}\times\mathrm{1}\;\mathbf{in}{}\<[E]%
\\
\>[B]{}\hsindent{14}{}\<[14]%
\>[14]{}\mathbf{let}\;\Varid{w}\mathrel{=}\Varid{u}\times(\Varid{v}\times\Varid{v})\;\mathbf{in}{}\<[E]%
\\
\>[B]{}\hsindent{14}{}\<[14]%
\>[14]{}\mathrm{1}\mathbin{/}(\Varid{w}\times\Varid{w}){}\<[27]%
\>[27]{}||\mskip1.5mu]{}\<[27E]%
\ColumnHook
\end{hscode}\resethooks
With the exception of the top-level term, all of the overhead of
lambda abstraction and function application has been removed; we
explain below why this is guaranteed by the subformula property.
From the normalised term it is easy to generate the final C code:
\begin{lstlisting}
float prog (float u) {
  float w; float v; float r;
  if (u == 0.0) {
    r = 0.0;
  } else {
    v = (u * 1.0);
    w = (u * (v * v));
    r = (1.0f / (w * w));
  }
  return r;
}
\end{lstlisting}
By default, we always generate a routine called \texttt{prog};
it is easy to provide the name as an additional parameter if required.



Depending on your point of view, quotation in this form of QDSL is
either desirable, because it makes manifest the staging, or
undesirable because it is too noisy.  QDSL enables us to ``steal'' the
entire syntax of the host language for our DSL.  In Haskell, an EDSL can
use the same syntax for arithmetic operators, but must use a different
syntax for equality tests and conditionals, as explained in
Section~\ref{sec:qdsl-vs-edsl}.

Within the quotation brackets there appear lambda abstractions and
function applications, while our intention is to generate first-order
code. How can the QDSL~Feldspar user be certain that such function
applications do not render transformation to first-order code
impossible or introduce additional runtime overhead?
The answer is the subformula property.

\subsection{The subformula property}
\label{subsec:subformula}

Gentzen's subformula property guarantees that any proof can be
normalised so that the only formulas that appear within it are
subformulas of one of the hypotheses or of the conclusion of the
proof.  Viewed through the lens of Propositions as Types,
Gentzen's subformula property guarantees that any term
can be normalised so that the type of each of its subterms is a
subformula of either the type of one of its free variables (corresponding
to hypotheses) or of the term itself (corresponding to the conclusion).
Here the subformulas of a type are the type itself and the subformulas of
its parts, where the parts of \ensuremath{\Varid{a}\to \Varid{b}} are \ensuremath{\Varid{a}} and \ensuremath{\Varid{b}}, the parts of
\ensuremath{(\Varid{a},\Varid{b})} are \ensuremath{\Varid{a}} and \ensuremath{\Varid{b}}, and types \ensuremath{\Conid{Int}} and \ensuremath{\Conid{Float}} have no
parts.
(See Proposition~\ref{prop:subformula}.)


Further, it is easy to adapt the original proof to guarantee a
sharpened subformula property: any term can be normalised so that the
type of each of its proper subterms is a proper subformula of either the
type of one of its free variables (corresponding to hypotheses) or the
term itself (corresponding to the conclusion).  Here the proper
subterms of a term are all subterms save for free variables and the
term itself, and the proper subformulas of a type are all subformulas
save for the type itself.  In the example of the previous subsection,
the sharpened subformula property guarantees that after normalisation
a closed term of type \ensuremath{\Varid{float}\to \Varid{float}} will only have proper subterms
of type \ensuremath{\Varid{float}}, which is indeed true for the normalised term.
(See Proposition~\ref{prop:sharpened}.)

The subformula property depends on normalisation of terms, but
complete normalisation is not always possible or desirable.  The
extent of normalisation may be controlled by introducing uninterpreted
constants.  In particular, we introduce the uninterpreted constant
\begin{hscode}\SaveRestoreHook
\column{B}{@{}>{\hspre}l<{\hspost}@{}}%
\column{E}{@{}>{\hspre}l<{\hspost}@{}}%
\>[B]{}\Varid{save}\mathbin{::}\Conid{Rep}\;\Varid{a}\Rightarrow \Varid{a}\to \Varid{a}{}\<[E]%
\ColumnHook
\end{hscode}\resethooks
of arity $1$, which is equivalent to the identity function
on representable types.  Unfolding of an application \ensuremath{\Conid{L}\;\Conid{M}}
can be inhibited by rewriting it in the form \ensuremath{\Varid{save}\;\Conid{L}\;\Conid{M}},
where \ensuremath{\Conid{L}} and \ensuremath{\Conid{M}} are arbitrary terms.
A use of \ensuremath{\Varid{save}} appears in Section~\ref{subsec:arrays}.
In a context with recursion, we take
\begin{hscode}\SaveRestoreHook
\column{B}{@{}>{\hspre}l<{\hspost}@{}}%
\column{E}{@{}>{\hspre}l<{\hspost}@{}}%
\>[B]{}\Varid{fix}\mathbin{::}(\Varid{a}\to \Varid{a})\to \Varid{a}{}\<[E]%
\ColumnHook
\end{hscode}\resethooks
as an uninterpreted constant.




\subsection{A second example}
\label{subsec:maybe}

In the previous code, we arbitrarily chose that raising zero to a
negative power yields zero. Say that we wish to exploit the \ensuremath{\Conid{Maybe}}
type of Haskell to refactor the code, by separating identification of the
exceptional case (negative exponent of zero) from choosing a value for
this case (zero).  We decompose \ensuremath{\Varid{power}} into two functions \ensuremath{\Varid{power'}}
and \ensuremath{\Varid{power''}}, where the first returns \ensuremath{\Conid{Nothing}} in the exceptional
case, and the second maps \ensuremath{\Conid{Nothing}} to a suitable value.

The \ensuremath{\Conid{Maybe}} type is a part of the Haskell standard prelude.
\begin{hscode}\SaveRestoreHook
\column{B}{@{}>{\hspre}l<{\hspost}@{}}%
\column{9}{@{}>{\hspre}c<{\hspost}@{}}%
\column{9E}{@{}l@{}}%
\column{13}{@{}>{\hspre}l<{\hspost}@{}}%
\column{E}{@{}>{\hspre}l<{\hspost}@{}}%
\>[B]{}\mathbf{data}\;\Conid{Maybe}\;\Varid{a}\mathrel{=}\Conid{Nothing}\mid \Conid{Just}\;\Varid{a}{}\<[E]%
\\
\>[B]{}\Varid{maybe}{}\<[9]%
\>[9]{}\mathbin{::}{}\<[9E]%
\>[13]{}\Varid{b}\to (\Varid{a}\to \Varid{b})\to \Conid{Maybe}\;\Varid{a}\to \Varid{b}{}\<[E]%
\\
\>[B]{}\Varid{return}{}\<[9]%
\>[9]{}\mathbin{::}{}\<[9E]%
\>[13]{}\Varid{a}\to \Conid{Maybe}\;\Varid{a}{}\<[E]%
\\
\>[B]{}(\bind ){}\<[9]%
\>[9]{}\mathbin{::}{}\<[9E]%
\>[13]{}\Conid{Maybe}\;\Varid{a}\to (\Varid{a}\to \Conid{Maybe}\;\Varid{b})\to \Conid{Maybe}\;\Varid{b}{}\<[E]%
\ColumnHook
\end{hscode}\resethooks

Here is the refactored code.
\begin{hscode}\SaveRestoreHook
\column{B}{@{}>{\hspre}l<{\hspost}@{}}%
\column{3}{@{}>{\hspre}l<{\hspost}@{}}%
\column{5}{@{}>{\hspre}l<{\hspost}@{}}%
\column{13}{@{}>{\hspre}l<{\hspost}@{}}%
\column{15}{@{}>{\hspre}l<{\hspost}@{}}%
\column{16}{@{}>{\hspre}l<{\hspost}@{}}%
\column{19}{@{}>{\hspre}l<{\hspost}@{}}%
\column{27}{@{}>{\hspre}l<{\hspost}@{}}%
\column{33}{@{}>{\hspre}l<{\hspost}@{}}%
\column{37}{@{}>{\hspre}l<{\hspost}@{}}%
\column{E}{@{}>{\hspre}l<{\hspost}@{}}%
\>[B]{}\Varid{power'}\mathbin{::}\Conid{Int}\to \Conid{Qt}\;(\Conid{Float}\to \Conid{Maybe}\;\Conid{Float}){}\<[E]%
\\
\>[B]{}\Varid{power'}\;\Varid{n}\mathrel{=}{}\<[E]%
\\
\>[B]{}\hsindent{3}{}\<[3]%
\>[3]{}\mathbf{if}\;\Varid{n}\mathbin{<}\mathrm{0}\;\mathbf{then}{}\<[E]%
\\
\>[3]{}\hsindent{2}{}\<[5]%
\>[5]{}[\mskip1.5mu ||\lambda \Varid{x}\to {}\<[16]%
\>[16]{}\mathbf{if}\;\Varid{x}\longeq \mathrm{0}\;{}\<[27]%
\>[27]{}\mathbf{then}\;{}\<[33]%
\>[33]{}\Conid{Nothing}{}\<[E]%
\\
\>[27]{}\mathbf{else}\;{}\<[33]%
\>[33]{}\mathbf{do}\;{}\<[37]%
\>[37]{}\Varid{y}\leftarrow \mathbin{\$\$}(\Varid{power'}\;(\mathbin{-}\Varid{n}))\;\Varid{x}{}\<[E]%
\\
\>[37]{}\Varid{return}\;(\mathrm{1}\mathbin{/}\Varid{y})||\mskip1.5mu]{}\<[E]%
\\
\>[B]{}\hsindent{3}{}\<[3]%
\>[3]{}\mathbf{else}\;\mathbf{if}\;\Varid{n}\longeq \mathrm{0}\;\mathbf{then}{}\<[E]%
\\
\>[3]{}\hsindent{2}{}\<[5]%
\>[5]{}[\mskip1.5mu ||\lambda \Varid{x}\to \Varid{return}\;\mathrm{1}||\mskip1.5mu]{}\<[E]%
\\
\>[B]{}\hsindent{3}{}\<[3]%
\>[3]{}\mathbf{else}\;\mathbf{if}\;\Varid{even}\;\Varid{n}\;\mathbf{then}{}\<[E]%
\\
\>[3]{}\hsindent{2}{}\<[5]%
\>[5]{}[\mskip1.5mu ||\lambda \Varid{x}\to \mathbf{do}\;{}\<[19]%
\>[19]{}\Varid{y}\leftarrow \mathbin{\$\$}(\Varid{power'}\;(\Varid{n}\rmdiv\mathrm{2}))\;\Varid{x}{}\<[E]%
\\
\>[19]{}\Varid{return}\;(\Varid{y}\times\Varid{y})||\mskip1.5mu]{}\<[E]%
\\
\>[B]{}\hsindent{3}{}\<[3]%
\>[3]{}\mathbf{else}{}\<[E]%
\\
\>[3]{}\hsindent{2}{}\<[5]%
\>[5]{}[\mskip1.5mu ||\lambda \Varid{x}\to \mathbf{do}\;{}\<[19]%
\>[19]{}\Varid{y}\leftarrow \mathbin{\$\$}(\Varid{power'}\;(\Varid{n}\mathbin{-}\mathrm{1}))\;\Varid{x}{}\<[E]%
\\
\>[19]{}\Varid{return}\;(\Varid{x}\times\Varid{y})||\mskip1.5mu]{}\<[E]%
\\[\blanklineskip]%
\>[B]{}\Varid{power''}\mathbin{::}{}\<[13]%
\>[13]{}\Conid{Int}\to \Conid{Qt}\;(\Conid{Float}\to \Conid{Float}){}\<[E]%
\\
\>[B]{}\Varid{power''}\;\Varid{n}\mathrel{=}{}\<[E]%
\\
\>[B]{}\hsindent{3}{}\<[3]%
\>[3]{}[\mskip1.5mu ||\lambda \Varid{x}\to {}\<[15]%
\>[15]{}\Varid{maybe}\;\mathrm{0}\;(\lambda \Varid{y}\to \Varid{y})\;(\mathbin{\$\$}(\Varid{power'}\;\Varid{n})\;\Varid{x})||\mskip1.5mu]{}\<[E]%
\ColumnHook
\end{hscode}\resethooks
Evaluation and normalisation of
\ensuremath{\Varid{power}\;(\mathbin{-}\mathrm{6})} and \ensuremath{\Varid{power''}\;(\mathbin{-}\mathrm{6})} yield identical terms
(up to renaming), and hence applying \ensuremath{\Varid{qdsl}} to these yields
identical C code.

The subformula property is key: because the final type of the result
does not involve \ensuremath{\Conid{Maybe}} it is certain that normalisation will remove
all its occurrences.  Occurrences of \ensuremath{\mathbf{do}} notation are expanded to
applications of \ensuremath{(\bind )}, as usual.  Rather than taking \ensuremath{\Varid{return}},
\ensuremath{(\bind )}, and \ensuremath{\Varid{maybe}} as uninterpreted constants (whose types have subformulas
involving \ensuremath{\Conid{Maybe}}), we treat them as known definitions to be
eliminated by the normaliser.  Type \ensuremath{\Conid{Maybe}} is a sum type,
and is normalised as described in Section~\ref{sec:subformula}.


\subsection{While}
\label{subsec:while}

Code that is intended to compile to a \text{\tt while} loop in C is indicated
in QDSL~Feldspar by application of \ensuremath{\Varid{while}}.
\begin{hscode}\SaveRestoreHook
\column{B}{@{}>{\hspre}l<{\hspost}@{}}%
\column{E}{@{}>{\hspre}l<{\hspost}@{}}%
\>[B]{}\Varid{while}\mathbin{::}(\Conid{Rep}\;\Varid{s})\Rightarrow (\Varid{s}\to \Conid{Bool})\to (\Varid{s}\to \Varid{s})\to \Varid{s}\to \Varid{s}{}\<[E]%
\ColumnHook
\end{hscode}\resethooks
Rather than using side-effects, \ensuremath{\Varid{while}} takes three
arguments: a predicate over the current state, of type \ensuremath{\Varid{s}\to \Conid{Bool}}; a
function from current state to new state, of type \ensuremath{\Varid{s}\to \Varid{s}}; and an
initial state of type \ensuremath{\Varid{s}}; and it returns a final state of type \ensuremath{\Varid{s}}.
So that we may compile \ensuremath{\Varid{while}} loops to C, the type of the state
is constrained to representable types.

We can define a \ensuremath{\Varid{for}} loop in terms of a \ensuremath{\Varid{while}} loop.
\begin{hscode}\SaveRestoreHook
\column{B}{@{}>{\hspre}l<{\hspost}@{}}%
\column{8}{@{}>{\hspre}l<{\hspost}@{}}%
\column{36}{@{}>{\hspre}l<{\hspost}@{}}%
\column{E}{@{}>{\hspre}l<{\hspost}@{}}%
\>[B]{}\Varid{for}\mathbin{::}(\Conid{Rep}\;\Varid{s})\Rightarrow \Conid{Qt}\;(\Conid{Int}\to \Varid{s}\to (\Conid{Int}\to \Varid{s}\to \Varid{s})\to \Varid{s}){}\<[E]%
\\
\>[B]{}\Varid{for}\mathrel{=}{}\<[8]%
\>[8]{}[\mskip1.5mu ||\lambda \Varid{n}\;\Varid{s}_{\mathrm{0}}\;\Varid{b}\to \Varid{snd}\;(\Varid{while}\;{}\<[36]%
\>[36]{}(\lambda (\Varid{i},\Varid{s})\to \Varid{i}\mathbin{<}\Varid{n})\;{}\<[E]%
\\
\>[36]{}(\lambda (\Varid{i},\Varid{s})\to (\Varid{i}\mathbin{+}\mathrm{1},\Varid{b}\;\Varid{i}\;\Varid{s}))\;{}\<[E]%
\\
\>[36]{}(\mathrm{0},\Varid{s}_{\mathrm{0}}))||\mskip1.5mu]{}\<[E]%
\ColumnHook
\end{hscode}\resethooks
The state of the \ensuremath{\Varid{while}} loop is a pair consisting of a counter and
the state of the \ensuremath{\Varid{for}} loop. The body \ensuremath{\Varid{b}} of the \ensuremath{\Varid{for}} loop is a function
that expects both the counter and the state of the \ensuremath{\Varid{for}} loop.
The counter is discarded when the loop is complete, and the final state
of the \ensuremath{\Varid{for}} loop returned. Here \ensuremath{\Varid{while}}, like \ensuremath{\Varid{snd}} and \ensuremath{(\mathbin{+})},
is a constant known to QDSL Feldspar, and so not enclosed in \ensuremath{\mathbin{\$\$}} antiquotes.

As an example, we can define Fibonacci using a \ensuremath{\Varid{for}} loop.
\begin{hscode}\SaveRestoreHook
\column{B}{@{}>{\hspre}l<{\hspost}@{}}%
\column{8}{@{}>{\hspre}l<{\hspost}@{}}%
\column{E}{@{}>{\hspre}l<{\hspost}@{}}%
\>[B]{}\Varid{fib}\mathbin{::}\Conid{Qt}\;(\Conid{Int}\to \Conid{Int}){}\<[E]%
\\
\>[B]{}\Varid{fib}\mathrel{=}{}\<[8]%
\>[8]{}[\mskip1.5mu ||\lambda \Varid{n}\to \Varid{fst}\;(\mathbin{\$\$}\Varid{for}\;\Varid{n}\;(\mathrm{0},\mathrm{1})\;(\lambda \Varid{i}\;(\Varid{a},\Varid{b})\to (\Varid{b},\Varid{a}\mathbin{+}\Varid{b})))||\mskip1.5mu]{}\<[E]%
\ColumnHook
\end{hscode}\resethooks

Again, the subformula property plays a key role.
As explained in Section~\ref{subsec:subformula}, primitives of the
language to be compiled, such as \ensuremath{(\times)} and \ensuremath{\Varid{while}}, are treated as
free variables or constants of a given arity.
As described in Section~\ref{sec:subformula},
we can ensure that after normalisation every occurence of \ensuremath{\Varid{while}}
has the form
\begin{hscode}\SaveRestoreHook
\column{B}{@{}>{\hspre}l<{\hspost}@{}}%
\column{E}{@{}>{\hspre}l<{\hspost}@{}}%
\>[B]{}\Varid{while}\;(\lambda \Varid{s}\to \cdots)\;(\lambda \Varid{s}\to \cdots)\;(\cdots){}\<[E]%
\ColumnHook
\end{hscode}\resethooks
where the first ellipses has type \ensuremath{\Conid{Bool}},
and both occurrences of the bound variable \ensuremath{\Varid{s}}
and the second and third ellipses all have the same type,
that of the state of the while loop.

Unsurprisingly, and in accord with the subformula property, each
occurrence of \ensuremath{\Varid{while}} in the normalised code will contain subterms
with the type of its state. The restriction of state to representable
types increases the utility of the subformula property. For instance,
since we have chosen that \ensuremath{\Conid{Maybe}} is not a representable type, we can
ensure that any top-level function without \ensuremath{\Conid{Maybe}} in its type will
normalise to code not containing \ensuremath{\Conid{Maybe}} in the type of any subterm.
In particular, \ensuremath{\Conid{Maybe}} cannot appear in the state of a \ensuremath{\Varid{while}} loop,
which is restricted to representable types.
An alternative choice is possible, as we will see in the next section.

\subsection{Arrays}
\label{subsec:arrays}

A key feature of Feldspar is its distinction between two types of
arrays, manifest arrays, \ensuremath{\Conid{Arr}}, which may appear at run-time, and
``pull arrays'', \ensuremath{\Conid{Vec}}, which are eliminated by fusion at generation-time.
Again, we exploit the subformula property to ensure
no subterms of type \ensuremath{\Conid{Vec}} remain in the final program.

The type \ensuremath{\Conid{Arr}} of manifest arrays is simply Haskell's array type,
specialised to arrays with integer indices and zero-based indexing.
The type \ensuremath{\Conid{Vec}} of pull arrays is defined in terms of existing types,
as a pair consisting of the length of the array and a function
that given an index returns the array element at that index.
\begin{hscode}\SaveRestoreHook
\column{B}{@{}>{\hspre}l<{\hspost}@{}}%
\column{13}{@{}>{\hspre}c<{\hspost}@{}}%
\column{13E}{@{}l@{}}%
\column{16}{@{}>{\hspre}l<{\hspost}@{}}%
\column{E}{@{}>{\hspre}l<{\hspost}@{}}%
\>[B]{}\mathbf{type}\;\Conid{Arr}\;\Varid{a}{}\<[13]%
\>[13]{}\mathrel{=}{}\<[13E]%
\>[16]{}\Conid{Array}\;\Conid{Int}\;\Varid{a}{}\<[E]%
\ColumnHook
\end{hscode}\resethooks
\begin{hscode}\SaveRestoreHook
\column{B}{@{}>{\hspre}l<{\hspost}@{}}%
\column{13}{@{}>{\hspre}c<{\hspost}@{}}%
\column{13E}{@{}l@{}}%
\column{16}{@{}>{\hspre}l<{\hspost}@{}}%
\column{E}{@{}>{\hspre}l<{\hspost}@{}}%
\>[B]{}\mathbf{data}\;\Conid{Vec}\;\Varid{a}{}\<[13]%
\>[13]{}\mathrel{=}{}\<[13E]%
\>[16]{}\Conid{Vec}\;\Conid{Int}\;(\Conid{Int}\to \Varid{a}){}\<[E]%
\ColumnHook
\end{hscode}\resethooks
Values of type \ensuremath{\Conid{Arr}} are representable, assuming that the
element type is representable, while values of type \ensuremath{\Conid{Vec}}
are not representable.
\begin{hscode}\SaveRestoreHook
\column{B}{@{}>{\hspre}l<{\hspost}@{}}%
\column{E}{@{}>{\hspre}l<{\hspost}@{}}%
\>[B]{}\mathbf{instance}\;(\Conid{Rep}\;\Varid{a})\Rightarrow \Conid{Rep}\;(\Conid{Arr}\;\Varid{a}){}\<[E]%
\ColumnHook
\end{hscode}\resethooks

For arrays, we assume the following primitive operations.
\begin{hscode}\SaveRestoreHook
\column{B}{@{}>{\hspre}l<{\hspost}@{}}%
\column{9}{@{}>{\hspre}c<{\hspost}@{}}%
\column{9E}{@{}l@{}}%
\column{13}{@{}>{\hspre}l<{\hspost}@{}}%
\column{E}{@{}>{\hspre}l<{\hspost}@{}}%
\>[B]{}\Varid{mkArr}{}\<[9]%
\>[9]{}\mathbin{::}{}\<[9E]%
\>[13]{}(\Conid{Rep}\;\Varid{a})\Rightarrow \Conid{Int}\to (\Conid{Int}\to \Varid{a})\to \Conid{Arr}\;\Varid{a}{}\<[E]%
\\
\>[B]{}\Varid{lnArr}{}\<[9]%
\>[9]{}\mathbin{::}{}\<[9E]%
\>[13]{}(\Conid{Rep}\;\Varid{a})\Rightarrow \Conid{Arr}\;\Varid{a}\to \Conid{Int}{}\<[E]%
\\
\>[B]{}\Varid{ixArr}{}\<[9]%
\>[9]{}\mathbin{::}{}\<[9E]%
\>[13]{}(\Conid{Rep}\;\Varid{a})\Rightarrow \Conid{Arr}\;\Varid{a}\to \Conid{Int}\to \Varid{a}{}\<[E]%
\ColumnHook
\end{hscode}\resethooks
The first populates a manifest array of the given
size using the given indexing function, the second
returns the length of the array, and the third returns
the array element at the given index.
Array components must be representable.

We define functions to convert between the two representations in the
obvious way.
\begin{hscode}\SaveRestoreHook
\column{B}{@{}>{\hspre}l<{\hspost}@{}}%
\column{14}{@{}>{\hspre}c<{\hspost}@{}}%
\column{14E}{@{}l@{}}%
\column{18}{@{}>{\hspre}l<{\hspost}@{}}%
\column{E}{@{}>{\hspre}l<{\hspost}@{}}%
\>[B]{}\Varid{toVec}{}\<[14]%
\>[14]{}\mathbin{::}{}\<[14E]%
\>[18]{}\Conid{Rep}\;\Varid{a}\Rightarrow \Conid{Qt}\;(\Conid{Arr}\;\Varid{a}\to \Conid{Vec}\;\Varid{a}){}\<[E]%
\\
\>[B]{}\Varid{toVec}{}\<[14]%
\>[14]{}\mathrel{=}{}\<[14E]%
\>[18]{}[\mskip1.5mu ||\lambda \Varid{a}\to \Conid{Vec}\;(\Varid{lnArr}\;\Varid{a})\;(\lambda \Varid{i}\to \Varid{ixArr}\;\Varid{a}\;\Varid{i})||\mskip1.5mu]{}\<[E]%
\\[\blanklineskip]%
\>[B]{}\Varid{fromVec}{}\<[14]%
\>[14]{}\mathbin{::}{}\<[14E]%
\>[18]{}\Conid{Rep}\;\Varid{a}\Rightarrow \Conid{Qt}\;(\Conid{Vec}\;\Varid{a}\to \Conid{Arr}\;\Varid{a}){}\<[E]%
\\
\>[B]{}\Varid{fromVec}{}\<[14]%
\>[14]{}\mathrel{=}{}\<[14E]%
\>[18]{}[\mskip1.5mu ||\lambda (\Conid{Vec}\;\Varid{n}\;\Varid{g})\to \Varid{mkArr}\;\Varid{n}\;\Varid{g}||\mskip1.5mu]{}\<[E]%
\ColumnHook
\end{hscode}\resethooks

It is straightforward to define operations on vectors,
including combining corresponding elements of two vectors,
summing the elements of a vector, dot product of two vectors,
and norm of a vector.
When combining two vectors, the length of the result is the
minimum of the lengths of the arguments.
\begin{hscode}\SaveRestoreHook
\column{B}{@{}>{\hspre}l<{\hspost}@{}}%
\column{10}{@{}>{\hspre}c<{\hspost}@{}}%
\column{10E}{@{}l@{}}%
\column{14}{@{}>{\hspre}l<{\hspost}@{}}%
\column{19}{@{}>{\hspre}l<{\hspost}@{}}%
\column{25}{@{}>{\hspre}l<{\hspost}@{}}%
\column{E}{@{}>{\hspre}l<{\hspost}@{}}%
\>[B]{}\Varid{minim}{}\<[10]%
\>[10]{}\mathbin{::}{}\<[10E]%
\>[14]{}\Conid{Ord}\;\Varid{a}\Rightarrow \Conid{Qt}\;(\Varid{a}\to \Varid{a}\to \Varid{a}){}\<[E]%
\\
\>[B]{}\Varid{minim}{}\<[10]%
\>[10]{}\mathrel{=}{}\<[10E]%
\>[14]{}[\mskip1.5mu ||\lambda \Varid{x}\;\Varid{y}\to \mathbf{if}\;\Varid{x}\mathbin{<}\Varid{y}\;\mathbf{then}\;\Varid{x}\;\mathbf{else}\;\Varid{y}||\mskip1.5mu]{}\<[E]%
\\[\blanklineskip]%
\>[B]{}\Varid{zipVec}{}\<[10]%
\>[10]{}\mathbin{::}{}\<[10E]%
\>[14]{}\Conid{Qt}\;((\Varid{a}\to \Varid{b}\to \Varid{c})\to \Conid{Vec}\;\Varid{a}\to \Conid{Vec}\;\Varid{b}\to \Conid{Vec}\;\Varid{c}){}\<[E]%
\\
\>[B]{}\Varid{zipVec}{}\<[10]%
\>[10]{}\mathrel{=}{}\<[10E]%
\>[14]{}[\mskip1.5mu ||{}\<[19]%
\>[19]{}\lambda \Varid{f}\;(\Conid{Vec}\;\Varid{m}\;\Varid{g})\;(\Conid{Vec}\;\Varid{n}\;\Varid{h})\to {}\<[E]%
\\
\>[19]{}\hsindent{6}{}\<[25]%
\>[25]{}\Conid{Vec}\;(\mathbin{\$\$}\Varid{minim}\;\Varid{m}\;\Varid{n})\;(\lambda \Varid{i}\to \Varid{f}\;(\Varid{g}\;\Varid{i})\;(\Varid{h}\;\Varid{i}))||\mskip1.5mu]{}\<[E]%
\\[\blanklineskip]%
\>[B]{}\Varid{sumVec}{}\<[10]%
\>[10]{}\mathbin{::}{}\<[10E]%
\>[14]{}(\Conid{Rep}\;\Varid{a},\Conid{Num}\;\Varid{a})\Rightarrow \Conid{Qt}\;(\Conid{Vec}\;\Varid{a}\to \Varid{a}){}\<[E]%
\\
\>[B]{}\Varid{sumVec}{}\<[10]%
\>[10]{}\mathrel{=}{}\<[10E]%
\>[14]{}[\mskip1.5mu ||\lambda (\Conid{Vec}\;\Varid{n}\;\Varid{g})\to \mathbin{\$\$}\Varid{for}\;\Varid{n}\;\mathrm{0}\;(\lambda \Varid{i}\;\Varid{x}\to \Varid{x}\mathbin{+}\Varid{g}\;\Varid{i})||\mskip1.5mu]{}\<[E]%
\\[\blanklineskip]%
\>[B]{}\Varid{dotVec}{}\<[10]%
\>[10]{}\mathbin{::}{}\<[10E]%
\>[14]{}(\Conid{Rep}\;\Varid{a},\Conid{Num}\;\Varid{a})\Rightarrow \Conid{Qt}\;(\Conid{Vec}\;\Varid{a}\to \Conid{Vec}\;\Varid{a}\to \Varid{a}){}\<[E]%
\\
\>[B]{}\Varid{dotVec}{}\<[10]%
\>[10]{}\mathrel{=}{}\<[10E]%
\>[14]{}[\mskip1.5mu ||\lambda \Varid{u}\;\Varid{v}\to \mathbin{\$\$}\Varid{sumVec}\;(\mathbin{\$\$}\Varid{zipVec}\;(\times)\;\Varid{u}\;\Varid{v})||\mskip1.5mu]{}\<[E]%
\\[\blanklineskip]%
\>[B]{}\Varid{normVec}{}\<[10]%
\>[10]{}\mathbin{::}{}\<[10E]%
\>[14]{}\Conid{Qt}\;(\Conid{Vec}\;\Conid{Float}\to \Conid{Float}){}\<[E]%
\\
\>[B]{}\Varid{normVec}{}\<[10]%
\>[10]{}\mathrel{=}{}\<[10E]%
\>[14]{}[\mskip1.5mu ||\lambda \Varid{v}\to \Varid{sqrt}\;(\mathbin{\$\$}\Varid{dotVec}\;\Varid{v}\;\Varid{v})||\mskip1.5mu]{}\<[E]%
\ColumnHook
\end{hscode}\resethooks
The third of these uses the \ensuremath{\Varid{for}} loop defined in
Section~\ref{subsec:while}.

Our final function cannot accept \ensuremath{\Conid{Vec}} as input, since
the \ensuremath{\Conid{Vec}} type is not representable, but it can accept
\ensuremath{\Conid{Arr}} as input.  For instance, if we invoke \ensuremath{\Varid{qdsl}} on
\[
\ensuremath{[\mskip1.5mu ||\mathbin{\$\$}\Varid{normVec}\hsdot{\circ }{.}\mathbin{\$\$}\Varid{toVec}||\mskip1.5mu]}
\]
the quoted term normalises to
\begin{hscode}\SaveRestoreHook
\column{B}{@{}>{\hspre}l<{\hspost}@{}}%
\column{54}{@{}>{\hspre}l<{\hspost}@{}}%
\column{61}{@{}>{\hspre}l<{\hspost}@{}}%
\column{62}{@{}>{\hspre}l<{\hspost}@{}}%
\column{66}{@{}>{\hspre}l<{\hspost}@{}}%
\column{E}{@{}>{\hspre}l<{\hspost}@{}}%
\>[B]{}[\mskip1.5mu ||\lambda \Varid{a}\to {}\<[61]%
\>[61]{}\Varid{sqrt}\;(\Varid{snd}{}\<[E]%
\\
\>[B]{}\hsindent{54}{}\<[54]%
\>[54]{}(\Varid{while}\;{}\<[62]%
\>[62]{}(\lambda \Varid{s}\to \Varid{fst}\;\Varid{s}\mathbin{<}\Varid{lnArr}\;\Varid{a})\;{}\<[E]%
\\
\>[62]{}(\lambda \Varid{s}\to \mathbf{let}\;\Varid{i}\mathrel{=}\Varid{fst}\;\Varid{s}\;\mathbf{in}{}\<[E]%
\\
\>[62]{}\hsindent{4}{}\<[66]%
\>[66]{}(\Varid{i}\mathbin{+}\mathrm{1},\Varid{snd}\;\Varid{s}\mathbin{+}(\Varid{ixArr}\;\Varid{a}\;\Varid{i}\times\Varid{ixArr}\;\Varid{a}\;\Varid{i})))\;{}\<[E]%
\\
\>[62]{}(\mathrm{0},\mathrm{0.0})))||\mskip1.5mu]{}\<[E]%
\ColumnHook
\end{hscode}\resethooks
from which it is easy to generate C code.

The vector representation makes it easy to define any
function where each vector element is computed independently,
such as the examples above,
vector append (\ensuremath{\Varid{appVec}})
and creating a vector of one element (\ensuremath{\Varid{uniVec}}),
but is less well suited to functions with dependencies
between elements, such as computing a running sum.

Types and the subformula property help us to guarantee fusion.  The
subformula property guarantees that all occurrences of \ensuremath{\Conid{Vec}} must be
eliminated, while occurrences of \ensuremath{\Conid{Arr}} will remain. There are some
situations where fusion is not beneficial, notably when an
intermediate vector is accessed many times, in which case fusion will
cause the elements to be recomputed. An alternative is to materialise
the vector as an array with the following function.
\begin{hscode}\SaveRestoreHook
\column{B}{@{}>{\hspre}l<{\hspost}@{}}%
\column{11}{@{}>{\hspre}c<{\hspost}@{}}%
\column{11E}{@{}l@{}}%
\column{15}{@{}>{\hspre}l<{\hspost}@{}}%
\column{E}{@{}>{\hspre}l<{\hspost}@{}}%
\>[B]{}\Varid{memorise}{}\<[11]%
\>[11]{}\mathbin{::}{}\<[11E]%
\>[15]{}\Conid{Rep}\;\Varid{a}\Rightarrow \Conid{Qt}\;(\Conid{Vec}\;\Varid{a}\to \Conid{Vec}\;\Varid{a}){}\<[E]%
\\
\>[B]{}\Varid{memorise}\mathrel{=}[\mskip1.5mu ||\mathbin{\$\$}\Varid{toVec}\hsdot{\circ }{.}\Varid{save}\hsdot{\circ }{.}\mathbin{\$\$}\Varid{fromVec}||\mskip1.5mu]{}\<[E]%
\ColumnHook
\end{hscode}\resethooks
Here we interpose \ensuremath{\Varid{save}}, as defined in Section~\ref{subsec:subformula}
to forestall the fusion that would otherwise occur. For example, if
\begin{hscode}\SaveRestoreHook
\column{B}{@{}>{\hspre}l<{\hspost}@{}}%
\column{28}{@{}>{\hspre}l<{\hspost}@{}}%
\column{E}{@{}>{\hspre}l<{\hspost}@{}}%
\>[B]{}\Varid{blur}\mathbin{::}\Conid{Qt}\;(\Conid{Vec}\;\Conid{Float}\to \Conid{Vec}\;\Conid{Float}){}\<[E]%
\\
\>[B]{}\Varid{blur}\mathrel{=}[\mskip1.5mu ||\lambda \Varid{a}\to \mathbin{\$\$}\Varid{zipVec}\;{}\<[28]%
\>[28]{}(\lambda \Varid{x}\;\Varid{y}\to \Varid{sqrt}\;(\Varid{x}\times\Varid{y}))\;{}\<[E]%
\\
\>[28]{}(\mathbin{\$\$}\Varid{appVec}\;(\mathbin{\$\$}\Varid{uniVec}\;\mathrm{0})\;\Varid{a})\;{}\<[E]%
\\
\>[28]{}(\mathbin{\$\$}\Varid{appVec}\;\Varid{a}\;(\mathbin{\$\$}\Varid{uniVec}\;\mathrm{0}))||\mskip1.5mu]{}\<[E]%
\ColumnHook
\end{hscode}\resethooks
computes the geometric mean of adjacent elements of a vector, then one may choose to
compute either
\begin{center}
\ensuremath{[\mskip1.5mu \mathbin{||\$\$}\Varid{blur}\hsdot{\circ }{.}\mathbin{\$\$}\Varid{blur}||\mskip1.5mu]} ~~~or~~~ \ensuremath{[\mskip1.5mu \mathbin{||\$\$}\Varid{blur}\hsdot{\circ }{.}\mathbin{\$\$}\Varid{memorise}\hsdot{\circ }{.}\mathbin{\$\$}\Varid{blur}||\mskip1.5mu]}
\end{center}
with different trade-offs between recomputation and memory use.
Strong guarantees for fusion in combination with \ensuremath{\Varid{memorise}} give the
programmer a simple interface which provides powerful optimisations
combined with fine control over memory usage.

We have described the application of the subformula to array
fusion as based on ``pull arrays'' \citep{svenningsson:combining},
but the same technique should also apply to other techniques that
support array fusion, such as ``push arrays'' \citep{claessen:push}.

\section{Implementation}
\label{sec:implementation}

The original EDSL~Feldspar generates values of a GADT
(called \ensuremath{\Conid{Dp}} in Section~\ref{sec:qdsl-vs-edsl}), with constructs
that represent \ensuremath{\Varid{while}} and manifest arrays similar to those
above. A backend then compiles values of type \ensuremath{\Conid{Dp}\;\Varid{a}} to C code.
QDSL~Feldspar provides a transformer from \ensuremath{\Conid{Qt}\;\Varid{a}} to \ensuremath{\Conid{Dp}\;\Varid{a}}, and
shares the EDSL~Feldspar backend.

The transformer from \ensuremath{\Conid{Qt}} to \ensuremath{\Conid{Dp}} performs the following steps.
\begin{itemize}
\item In any context where a constant $c$ is not fully applied,
  it replaces $c$ with $\lambda \overline{x}.\, c \app \overline{x}$.
  It replaces identifiers connected to the type \ensuremath{\Conid{Maybe}}, such as
  \ensuremath{\Varid{return}}, \ensuremath{(\bind )}, and \ensuremath{\Varid{maybe}}, by their definitions.
\item It normalises the term to ensure the subformula property, using
  the rules of Section~\ref{sec:subformula}. The normaliser supports
  a limited set of types, including tuples, \ensuremath{\Conid{Maybe}}, and \ensuremath{\Conid{Vec}}.
\item It performs simple type inference, which is used to resolve
  overloading. Overloading is limited to a fixed set of
  cases, including overloading arithmetic operators at types
  \ensuremath{\Conid{Int}} and \ensuremath{\Conid{Float}}.
\item It traverses the term, converting \ensuremath{\Conid{Qt}} to \ensuremath{\Conid{Dp}}.
  It checks that only permitted primitives appear in \ensuremath{\Conid{Qt}},
  and translates these to their corresponding representation
  in \ensuremath{\Conid{Dp}}. Permitted primitives include:
  \ensuremath{(\longeq )}, \ensuremath{(\mathbin{<})}, \ensuremath{(\mathbin{+})}, \ensuremath{(\times)}, and similar, plus
  \ensuremath{\Varid{while}}, \ensuremath{\Varid{makeArr}}, \ensuremath{\Varid{lenArr}}, \ensuremath{\Varid{ixArr}}, and \ensuremath{\Varid{save}}.
\end{itemize}

An unfortunate feature of typed quasiquotation in GHC is that the
implementation discards all type information when creating the
representation of a term.  Type \ensuremath{\Conid{Qt}\;\Varid{a}} is a synonym for the type
\[
\ensuremath{\Conid{\Conid{TH}.Q}\;(\Conid{\Conid{TH}.TExp}\;\Varid{a})}
\]
where \ensuremath{\Conid{TH}} denotes the library for Template Haskell, \ensuremath{\Conid{\Conid{TH}.Q}} is the
quotation monad (used to look up identifiers and
generate fresh names), and \ensuremath{\Conid{\Conid{TH}.TExp}\;\Varid{a}} is the parse tree for a quoted
expression returning a value of type \ensuremath{\Varid{a}} (a wrapper for the type
\ensuremath{\Conid{\Conid{TH}.Exp}} of untyped expressions, with \ensuremath{\Varid{a}} as a phantom variable).
Thus, the
translator from \ensuremath{\Conid{Qt}\;\Varid{a}} to \ensuremath{\Conid{Dp}\;\Varid{a}} is forced to re-infer all type for
subterms, and for this reason we support only limited overloading, and we
translate the \ensuremath{\Conid{Maybe}} monad as a special case rather than supporting
overloading for monad operations in general.

The backend performs three transformations over \ensuremath{\Conid{Dp}} terms
before compiling to C. First, common subexpessions are recognised
and transformed to \ensuremath{\mathbf{let}} bindings.
Second, \ensuremath{\Conid{Dp}} terms are normalised using exactly the same rules
used for normalising \ensuremath{\Conid{Qt}} terms, as described in Section~\ref{sec:subformula}.
Third, \ensuremath{\Conid{Dp}} terms are optimised using $\eta$ contraction for
conditionals and arrays:
\[
\begin{array}{rcl}
          \ensuremath{\mathbf{if}\;\Conid{L}\;\mathbf{then}\;\Conid{M}\;\mathbf{else}\;\Conid{M}} & \rewrite{} & M \\
\ensuremath{\Varid{makeArr}\;(\Varid{lenArr}\;\Conid{M})\;(\Varid{ixArr}\;\Conid{M})} & \rewrite{} & M
\end{array}
\]
and a restricted form of linear inlining for \ensuremath{\mathbf{let}} bindings
that preserves the order of evaluation.

\newcommand{\ct}{\:Compile\:}
\newcommand{\rt}{\:Run\:}

\begin{figure}
Lines of Haskell code
\begin{center}
\begin{tabular}{|l|r|r|r|}
\hline
 & \makebox[22pt][c]{shared}
 & \makebox[22pt][c]{unique}
 & \makebox[22pt][c]{total}
\\ \hline
QDSL Feldspar & 3970 & 1722 & 5962 \\
EDSL Feldspar & 3970 &  452 & 4422 \\
\hline
\end{tabular}
\end{center}

Benchmarks
\begin{center}
\begin{tabular}{|l|l|}
\hline
IPGray     & Image Processing (Grayscale)  \\
IPBW       & Image Processing (Black and White) \\
FFT        & Fast Fourier Transform \\
CRC        & Cyclic Redundancy Check \\
Window     & Average array in a sliding window \\
\hline
\end{tabular}
\end{center}


Performance
\begin{center}
\begin{tabular}{|@{\:}l@{\:}|@{}c@{}|@{}c@{}|@{}c@{}|@{}c@{}|@{}c@{}|@{}c@{}|}
\hline
 & \multicolumn{2}{@{\:}c@{\:}}{QDSL Feldspar}
 & \multicolumn{2}{|@{\:}c@{\:}}{EDSL Feldspar}
 & \multicolumn{2}{|@{\:}c@{\:}|}{Generated Code}
\\ \hline
 & \ct & \rt & \ct & \rt & \ct & \rt
\\ \hline
IPGray & 16.96 & 0.01 & 15.06 & 0.01 & 0.06 & 0.39 \\
IPBW   & 17.08 & 0.01 & 14.86 & 0.01 & 0.06 & 0.19 \\
FFT    & 17.87 & 0.39 & 15.79 & 0.09 & 0.07 & 3.02 \\
CRC    & 17.14 & 0.01 & 15.33 & 0.01 & 0.05 & 0.12 \\
Window & 17.85 & 0.02 & 15.77 & 0.01 & 0.06 & 0.27
\\ \hline
\end{tabular}
\end{center}
Times in seconds; minimum time of ten runs. \\
Quad-core Intel i7-2640M CPU, 2.80 GHz, 3.7 GiB RAM.\\
GHC 7.8.3; GCC 4.8.2; Ubuntu 14.04 (64-bit).

\caption{Comparison of QDSL and EDSL Feldspar}
\label{fig:thetable}
\end{figure}

Figure~\ref{fig:thetable} lists lines of code,
benchmarks used,
and performance results.
The translator from \ensuremath{\Conid{Dp}} to C is shared by QDSL and EDSL Feldspar,
and listed in a separate column.
All five benchmarks run under QDSL and EDSL Felsdpar generate
identical C code, up to permutation of independent assignments, with
identical compile and run times.
The columns for QDSL and EDSL Feldspar give compile and run
times for Haskell, while the columns for generated code
give compile and run times for the generated C.
QDSL compile times are slightly greater than EDSL,
and QDSL run times range from identical to four times that of EDSL,
the increase being due to normalisation time
(our normaliser was not designed to be particularly efficient).

\section{The subformula property}
\label{sec:subformula}

This section introduces reduction rules for normalising terms that
enforce the subformula property while preserving sharing. The rules
adapt to both call-by-need and call-by-value.
We work with simple types. The only polymorphism in our examples
corresponds to instantiating constants (such as $\mathit{while}$) at
different types.

Types, terms, and values are presented in Figure~\ref{fig:term}.
Let $A$, $B$, $C$ range over types, including base types ($\iota$),
functions ($A \to B$), products ($A \times B$), and sums ($A + B$).
Let $L$, $M$, $N$ range over terms, and $x$, $y$, $z$ range over
variables.  Let $c$ range over constants, which are fully
applied according to their arity, as discussed below.
As constant applications are non-values, we represent literals as free
variables.
As usual, terms are taken as
equivalent up to renaming of bound variables. Write $\fv{M}$ for
the set of free variables of $M$, and $\subst{N}{x}{M}$ for
capture-avoiding substitution of $M$ for $x$ in $N$.
Let $V$, $W$ range over values.

Let $\Gamma$ range over type environments, which pair
variables with types, and write $\Gamma \vdash M:A$ to
indicate that term $M$ has type $A$ under type environment
$\Gamma$. Typing rules are standard.

Reduction rules for normalisation are presented in Figure~\ref{fig:norm}.
The rules are confluent, so order of application is irrelevant to the
final answer, but we break them into three phases to ease the proof
of strong normalisation. It is easy to confirm that all
of the reduction rules preserve sharing and preserve order of evaluation.

Write $M \mapsto_i N$ to indicate that $M$ reduces to $N$ in phase
$i$.
Let $F$ and $G$ range over two different forms of evaluation
frame used in Phases~1 and~2 respectively. Write $\fv{F}$ for the
set of free variables of $F$, and similarly for $G$.
Reductions are closed under compatible closure.

\figterm
\fignorm

The normalisation procedure consists of exhaustively applying the
reductions of Phase~1 until no more apply, then similarly for Phase~2,
and finally for Phase~3.  Phase~1 performs let-insertion, naming
subterms, along the lines of a translation to
A-normal form \citep{a-normal-form} or reductions (let.1) and (let.2)
in Moggi's metalanguage for monads \citep{Moggi-1991}.
Phase~2 performs two kinds of reduction: $\beta$ rules apply when an
introduction (construction) is immediately followed by an elimination
(deconstruction), and $\kappa$ rules push eliminators closer to
introducers to enable $\beta$ rules.  Phase~3 ``garbage collects''
unused terms as in the call-by-need lambda calculus
\citep{call-by-need}. Phase~3 should be omitted if the intended
semantics of the target language is call-by-value rather than
call-by-need.

Every term has a normal form.
\begin{proposition}[Strong normalisation]
Each of the reduction relations $\rewrite{i}$ is confluent and strongly
normalising: all $\rewrite{i}$ reduction sequences on well-typed
terms are finite.
\end{proposition}
The only non-trivial proof is for $\rewrite{2}$, which can be proved
via a standard reducibility argument (see, for example,
\cite{Lindley07}). If the target language includes general recursion,
normalisation should treat the fixpoint operator as an uninterpreted
constant.


The \emph{subformulas} of a type are the type itself and its
components. For instance, the subformulas of $A \to B$ are itself and
the subformulas of $A$ and $B$. The \emph{proper subformulas} of a
type are all its subformulas other than the type itself.

The \emph{subterms} of term are the term itself and its components.
For instance, the subterms of $\expabs{x}{}{N}$ are itself and the
subterms of $N$ and the subterms of $\expapp{L}{M}$ are itself and the
subterms of $L$ and $M$. The \emph{proper subterms} of a
term are all its subterms other than the term itself.

Constants are always fully applied; they are introduced as a
separate construct to avoid consideration of irrelevant subformulas
and subterms.
The type of a constant $c$ of arity $k$ is written
\[
c : A_1 \to \cdots \to A_k \to B
\]
and its subformulas are itself and $A_1$, \ldots, $A_k$, and $B$
(but not $A_i \to \cdots \to A_k \to B$ for $i > 1$).
An application of a constant $c$ of arity $k$ is written
\[
c \app M_1 \app \cdots \app M_k
\]
and its subterms are itself and $M_1$, \ldots, $M_k$
(but not $c \app M_1 \app \cdots \app M_j$ for $j < k$).
Free variables are equivalent to constants of arity zero.

Terms in normal form satisfy the subformula property.
\begin{proposition}[Subformula property]
\label{prop:subformula}
If $\Gamma \vdash M:A$ and $M$ is in normal form,
then every subterm of $M$ has a type that is either a subformula of $A$,
a subformula of a type in $\Gamma$, or a subformula of the type
of a constant in $M$.
\end{proposition}
The proof follows the lines of \citet{Prawitz-1965}.
The differences are that we have introduced fully applied constants
(to enable the sharpened subformula property, below), and that our
reduction rules introduce \ensuremath{\mathbf{let}}, in order to ensure sharing is preserved.

Normalisation may lead to an exponential or worse blow up in the size
of a term, for instance when there are nested \ensuremath{\mathbf{case}} expressions.
The benchmarks in Section~\ref{sec:implementation} do not suffer
from blow up, but it may be a problem in some contexts.
Normalisation may be controlled by introduction of
uninterpreted constants, as in Section~\ref{subsec:subformula}.
Further work is needed to understand when
complete normalisation is
desirable and when it is problematic.

Examination of the proof in \citet{Prawitz-1965} shows that in fact
normalisation achieves a sharper property.
\begin{proposition}[Sharpened subformula]
\label{prop:sharpened}
If $\Gamma \vdash M:A$ and $M$ is in normal form, then every proper
subterm of $M$ that is not a free variable or a subterm of a constant
application has a type that is a proper subformula of $A$ or a proper
subformula of a type in $\Gamma$.
\end{proposition}
We believe we are the first to formulate the sharpened version.

The sharpened subformula property says nothing about the types of
subterms of constant applications, but such types are immediately apparent by
recursive application of the sharpened subformula property.  Given a
subterm that is a constant application $c \app \overline{M}$, where
$c$ has type $\overline{A} \to B$, then the subterm itself has type
$B$, each subterm $M_i$ has type $A_i$, and every proper subterm of
$M_i$ that is not a free variable of $M_i$ or a subterm of a constant
application has a type that is a proper subformula of $A_i$ or a
proper subformula of the type of one of its free variables.

In Section~\ref{sec:qfeldspar}, we require that every
top-level term passed to \ensuremath{\Varid{qdsl}} is suitable for translation to C after
normalisation, and any DSL translating to
a \emph{first-order} language must impose a similar requirement.
One might at first guess the required property is that every
subterm is \emph{representable}, in the sense introduced in
Section~\ref{subsec:top}, but this is not quite right. The top-level
term is a function from a representable type to a representable type,
and the constant \ensuremath{\Varid{while}} expects subterms of type \ensuremath{\Varid{s}\to \Conid{Bool}}
and \ensuremath{\Varid{s}\to \Varid{s}}, where the state \ensuremath{\Varid{s}} is representable.  Fortunately,
the property required is not hard to formulate in a general way,
and is easy to ensure by applying the sharpened subformula property.

Take the
representable types to be any set closed under subformulas
that does not include function types.
We introduce a variant of the usual notion of \emph{rank} of a type,
with respect to a notion of representability.  A term of type \ensuremath{\Conid{A}\to \Conid{B}}
has rank $\min(m+1,n)$ where $m$ is the rank of \ensuremath{\Conid{A}} and $n$ is the
rank of \ensuremath{\Conid{B}}, while a term of representable type has rank $0$.
We say a term is \emph{first-order} when every subterm is either
representable, or is of the form $\lambda \overline{x}.\, N$
where each bound variable and the body is of representable type.

The following characterises translation to a first-order language.
\vspace{-2ex}
\begin{proposition}[First-order]
\label{prop:fo}
Consider a term of rank $1$, where every free variable has
rank $0$ and every constant has rank at most $2$.  Then the term
normalises to a term that is first-order.
\end{proposition}
The property follows immediately by observing that any term $L$
of rank $1$ can be rewritten to the form
$\lambda \overline{y}.\, (L \app \overline{y})$
where each bound variable
and the body has representable type, and then normalising and applying
the sharpened subformula property.

In QDSL Feldspar, \ensuremath{\Varid{while}} is a constant with type of rank $2$ and
other constants have types of rank $1$.  Section~\ref{subsec:arrays}
gives an example of a normalised term.  By the proposition, each
subterm has a representable type (boolean, integer, float, or a pair of an
integer and float) or is a lambda abstraction with bound variables and
body of representable type; and it is this property which ensures it
is easy to generate C code from the term.

\section{Feldspar as an EDSL}
\label{sec:qdsl-vs-edsl}

This section reviews the combination of deep and shallow embeddings
required to implement Feldspar as an EDSL, and considers
the trade-offs between the QDSL and EDSL approaches.  Much of this
section reprises \citet{svenningsson:combining}.

The top-level function of EDSL Feldspar has the type:
\begin{hscode}\SaveRestoreHook
\column{B}{@{}>{\hspre}l<{\hspost}@{}}%
\column{E}{@{}>{\hspre}l<{\hspost}@{}}%
\>[B]{}\Varid{edsl}\mathbin{::}(\Conid{Rep}\;\Varid{a},\Conid{Rep}\;\Varid{b})\Rightarrow (\Conid{Dp}\;\Varid{a}\to \Conid{Dp}\;\Varid{b})\to \Conid{C}{}\<[E]%
\ColumnHook
\end{hscode}\resethooks
Here \ensuremath{\Conid{Dp}\;\Varid{a}} is the deep representation of a term of type \ensuremath{\Varid{a}}.
The deep representation is described in detail in Section~\ref{subsec:deep}
below, and is chosen to be easy to translate to C.
As before, type \ensuremath{\Conid{C}} represents code in C,
and type class \ensuremath{\Conid{Rep}} restricts to representable types.


\subsection{A first example}
\label{subsec:e-power}

Here is the power function of Section~\ref{subsec:power},
written as an EDSL:
\begin{hscode}\SaveRestoreHook
\column{B}{@{}>{\hspre}l<{\hspost}@{}}%
\column{3}{@{}>{\hspre}l<{\hspost}@{}}%
\column{5}{@{}>{\hspre}l<{\hspost}@{}}%
\column{12}{@{}>{\hspre}c<{\hspost}@{}}%
\column{12E}{@{}l@{}}%
\column{21}{@{}>{\hspre}l<{\hspost}@{}}%
\column{E}{@{}>{\hspre}l<{\hspost}@{}}%
\>[B]{}\Varid{power}\mathbin{::}\Conid{Int}\to \Conid{Dp}\;\Conid{Float}\to \Conid{Dp}\;\Conid{Float}{}\<[E]%
\\
\>[B]{}\Varid{power}\;\Varid{n}\;\Varid{x}{}\<[12]%
\>[12]{}\mathrel{=}{}\<[12E]%
\\
\>[B]{}\hsindent{3}{}\<[3]%
\>[3]{}\mathbf{if}\;\Varid{n}\mathbin{<}\mathrm{0}\;\mathbf{then}{}\<[E]%
\\
\>[3]{}\hsindent{2}{}\<[5]%
\>[5]{}\Varid{x}\mathbin{{.}{\longeq }{.}}\mathrm{0}\mathbin{?}(\mathrm{0},{}\<[21]%
\>[21]{}\mathrm{1}\mathbin{/}\Varid{power}\;(\mathbin{-}\Varid{n})\;\Varid{x}){}\<[E]%
\\
\>[B]{}\hsindent{3}{}\<[3]%
\>[3]{}\mathbf{else}\;\mathbf{if}\;\Varid{n}\longeq \mathrm{0}\;\mathbf{then}{}\<[E]%
\\
\>[3]{}\hsindent{2}{}\<[5]%
\>[5]{}\mathrm{1}{}\<[E]%
\\
\>[B]{}\hsindent{3}{}\<[3]%
\>[3]{}\mathbf{else}\;\mathbf{if}\;\Varid{even}\;\Varid{n}\;\mathbf{then}{}\<[E]%
\\
\>[3]{}\hsindent{2}{}\<[5]%
\>[5]{}\mathbf{let}\;\Varid{y}\mathrel{=}\Varid{power}\;(\Varid{n}\rmdiv\mathrm{2})\;\Varid{x}\;\mathbf{in}\;\Varid{y}\times\Varid{y}{}\<[E]%
\\
\>[B]{}\hsindent{3}{}\<[3]%
\>[3]{}\mathbf{else}{}\<[E]%
\\
\>[3]{}\hsindent{2}{}\<[5]%
\>[5]{}\Varid{x}\times\Varid{power}\;(\Varid{n}\mathbin{-}\mathrm{1})\;\Varid{x}{}\<[E]%
\ColumnHook
\end{hscode}\resethooks
Type \ensuremath{\Conid{Q}\;(\Conid{Float}\to \Conid{Float})} in the QDSL variant becomes the type
\ensuremath{\Conid{Dp}\;\Conid{Float}\to \Conid{Dp}\;\Conid{Float}} in the EDSL variant, meaning that \ensuremath{\Varid{power}\;\Varid{n}}
accepts a representation
of the argument and returns a representation of that argument raised
to the $n$'th power.

In the EDSL variant, no quotation is required, and the code looks
almost---but not quite!---like an unstaged version of power, but with
different types.  Clever encoding tricks, explained later, permit
declarations, function calls, arithmetic operations, and numbers to
appear the same whether they are to be executed at generation-time or
run-time.  However, as explained later, comparison and conditionals
appear differently depending on whether they are to be executed at
generation-time or run-time, using \ensuremath{\Conid{M}\longeq \Conid{N}} and \ensuremath{\mathbf{if}\;\Conid{L}\;\mathbf{then}\;\Conid{M}\;\mathbf{else}\;\Conid{N}}
for the former but \ensuremath{\Conid{M}\mathbin{{.}{\longeq }{.}}\Conid{N}} and \ensuremath{\Conid{L}\mathbin{?}(\Conid{M},\Conid{N})} for the latter.

Invoking \ensuremath{\Varid{edsl}\;(\Varid{power}\;(\mathbin{-}\mathrm{6}))} generates code to raise a number to its \ensuremath{\mathbin{-}\mathrm{6}} power.
Evaluating \ensuremath{\Varid{power}\;(\mathbin{-}\mathrm{6})\;\Varid{u}}, where \ensuremath{\Varid{u}} is a term representing a variable of type \ensuremath{\Conid{Dp}\;\Conid{Float}},
yields the following:
\begin{hscode}\SaveRestoreHook
\column{B}{@{}>{\hspre}l<{\hspost}@{}}%
\column{3}{@{}>{\hspre}l<{\hspost}@{}}%
\column{10}{@{}>{\hspre}l<{\hspost}@{}}%
\column{E}{@{}>{\hspre}l<{\hspost}@{}}%
\>[B]{}(\Varid{u}\mathbin{{.}{\longeq }{.}}\mathrm{0})\mathbin{?}(\mathrm{0},{}\<[E]%
\\
\>[B]{}\hsindent{3}{}\<[3]%
\>[3]{}\mathrm{1}\mathbin{/}({}\<[10]%
\>[10]{}(\Varid{u}\times((\Varid{u}\times\mathrm{1})\times(\Varid{u}\times\mathrm{1})))\times{}\<[E]%
\\
\>[10]{}(\Varid{u}\times((\Varid{u}\times\mathrm{1})\times(\Varid{u}\times\mathrm{1}))))){}\<[E]%
\ColumnHook
\end{hscode}\resethooks
Applying common-subexpression elimination
permits recovering the sharing structure.
\[
\begin{array}{c@{~~}|@{~~}l}
\ensuremath{\Varid{v}} & \ensuremath{(\Varid{u}\times\mathrm{1})}  \\
\ensuremath{\Varid{w}} & \ensuremath{\Varid{u}\times(\Varid{v}\times\Varid{v})} \\
\text{top} & \ensuremath{(\Varid{u}\mathbin{{.}{\longeq }{.}}\mathrm{0})\mathbin{?}(\mathrm{0},\mathrm{1}\mathbin{/}(\Varid{w}\times\Varid{w}))}
\end{array}
\]
From the above, it is easy to generate the final C code,
which is identical to that in Section~\ref{subsec:power}.

Here are points of comparison between the two approaches.
\begin{itemize}

\item A function \ensuremath{\Varid{a}\to \Varid{b}} is embedded in
QDSL as \ensuremath{\Conid{Qt}\;(\Varid{a}\to \Varid{b})}, a representation of a function, and in
EDSL as \ensuremath{\Conid{Dp}\;\Varid{a}\to \Conid{Dp}\;\Varid{b}}, a function between representations.

\item QDSL enables the host and embedded languages to appear
identical.  In contrast, in Haskell, EDSL requires some term forms,
such as comparison and conditionals, to differ between the host and
embedded languages.  Other languages, notably Scala Virtualised
\citep{rompf2013scala}, may support more general overloading that
allows even comparison and conditionals to be identical.

\item QDSL requires syntax to separate quoted and unquoted terms. In
contrast, EDSL permits the host and embedded languages to intermingle
seamlessly. Depending on your point of view, explicit quotation syntax
may be considered as an unnecessary distraction or as drawing a useful
distinction between generation-time and run-time.  If one takes the
former view, the type-based approach to quotation found in C\# and
Scala might be preferred.

\item QDSL may share the same representation for quoted terms across a
range of applications; the quoted language is the host language, and
does not vary with the specific domain.  In contrast, EDSL typically
develops custom shallow and deep embeddings for each application;
a notable exception is the LMS and Delite frameworks for Scala,
which provide a deep embedding shared across several disparate DSLs
\citep{sujeeth2013composition}.

\item QDSL yields an unwieldy term that requires normalisation.  In
contrast, EDSL yields the term in normalised form in this case, though
there are other situations where a normaliser is required (see
Section~\ref{subsec:e-maybe}).

\item QDSL requires traversing the quoted term to ensure it only
mentions permitted identifiers. In contrast, EDSL guarantees that if a
term has the right type it will translate to the target.  If the
requirement to eyeball code to ensure only permitted identifiers are
used is considered too onerous, it should be easy to build a
preprocessor that checks this property. For example, in Haskell, it is
possible to incorporate such a preprocessor using MetaHaskell
\cite{metahaskell}.

\item Since QDSLs may share the same quoted terms across a range of
applications, the cost of building a normaliser or a preprocessor
might be amortised across multiple QDSLs for a single language.  In
the conclusion, we consider the design of a tool for building QDSLs
that uses a shared normaliser and preprocessor.


\item Once the deep embedding or the normalised quoted term is
produced, generating the domain-specific code is similar for both
approaches.

\end{itemize}

\subsection{A second example}
\label{subsec:e-maybe}

In Section~\ref{subsec:maybe}, we exploited the \ensuremath{\Conid{Maybe}} type to refactor the code.

In EDSL, we must use a new type, where
\ensuremath{\Conid{Maybe}}, \ensuremath{\Conid{Nothing}}, \ensuremath{\Conid{Just}}, and \ensuremath{\Varid{maybe}} become
\ensuremath{\Conid{Opt}}, \ensuremath{\Varid{none}}, \ensuremath{\Varid{some}}, and \ensuremath{\Varid{option}},
and \ensuremath{\Varid{return}} and \ensuremath{(\bind )} are similar to before.
\begin{hscode}\SaveRestoreHook
\column{B}{@{}>{\hspre}l<{\hspost}@{}}%
\column{9}{@{}>{\hspre}c<{\hspost}@{}}%
\column{9E}{@{}l@{}}%
\column{13}{@{}>{\hspre}l<{\hspost}@{}}%
\column{15}{@{}>{\hspre}l<{\hspost}@{}}%
\column{E}{@{}>{\hspre}l<{\hspost}@{}}%
\>[B]{}\mathbf{type}\;\Conid{Opt}\;\Varid{a}{}\<[E]%
\\
\>[B]{}\Varid{none}{}\<[9]%
\>[9]{}\mathbin{::}{}\<[9E]%
\>[13]{}\Conid{Undef}\;\Varid{a}\Rightarrow \Conid{Opt}\;\Varid{a}{}\<[E]%
\\
\>[B]{}\Varid{some}{}\<[9]%
\>[9]{}\mathbin{::}{}\<[9E]%
\>[13]{}\Varid{a}\to \Conid{Opt}\;\Varid{a}{}\<[E]%
\\
\>[B]{}\Varid{return}{}\<[9]%
\>[9]{}\mathbin{::}{}\<[9E]%
\>[13]{}\Varid{a}\to \Conid{Opt}\;\Varid{a}{}\<[E]%
\\
\>[B]{}(\bind ){}\<[9]%
\>[9]{}\mathbin{::}{}\<[9E]%
\>[13]{}\Conid{Opt}\;\Varid{a}\to (\Varid{a}\to \Conid{Opt}\;\Varid{b})\to \Conid{Opt}\;\Varid{b}{}\<[E]%
\\
\>[B]{}\Varid{option}{}\<[9]%
\>[9]{}\mathbin{::}{}\<[9E]%
\>[13]{}(\Conid{Undef}\;\Varid{a},\Conid{Undef}\;\Varid{b})\Rightarrow {}\<[E]%
\\
\>[13]{}\hsindent{2}{}\<[15]%
\>[15]{}\Varid{b}\to (\Varid{a}\to \Varid{b})\to \Conid{Opt}\;\Varid{a}\to \Varid{b}{}\<[E]%
\ColumnHook
\end{hscode}\resethooks
Type class \ensuremath{\Conid{Undef}} is explained in Section~\ref{subsec:undef},
and details of type \ensuremath{\Conid{Opt}} are given in Section~\ref{subsec:opt}.


Here is the refactored code.
\begin{hscode}\SaveRestoreHook
\column{B}{@{}>{\hspre}l<{\hspost}@{}}%
\column{3}{@{}>{\hspre}l<{\hspost}@{}}%
\column{5}{@{}>{\hspre}l<{\hspost}@{}}%
\column{9}{@{}>{\hspre}l<{\hspost}@{}}%
\column{13}{@{}>{\hspre}l<{\hspost}@{}}%
\column{14}{@{}>{\hspre}c<{\hspost}@{}}%
\column{14E}{@{}l@{}}%
\column{17}{@{}>{\hspre}l<{\hspost}@{}}%
\column{21}{@{}>{\hspre}l<{\hspost}@{}}%
\column{25}{@{}>{\hspre}l<{\hspost}@{}}%
\column{E}{@{}>{\hspre}l<{\hspost}@{}}%
\>[B]{}\Varid{power'}\mathbin{::}\Conid{Int}\to \Conid{Dp}\;\Conid{Float}\to \Conid{Opt}\;(\Conid{Dp}\;\Conid{Float}){}\<[E]%
\\
\>[B]{}\Varid{power'}\;\Varid{n}\;\Varid{x}\mathrel{=}{}\<[E]%
\\
\>[B]{}\hsindent{3}{}\<[3]%
\>[3]{}\mathbf{if}\;\Varid{n}\mathbin{<}\mathrm{0}\;\mathbf{then}{}\<[E]%
\\
\>[3]{}\hsindent{2}{}\<[5]%
\>[5]{}(\Varid{x}\mathbin{{.}{\longeq }{.}}\mathrm{0})\mathbin{?}({}\<[21]%
\>[21]{}\Varid{none},{}\<[E]%
\\
\>[21]{}\mathbf{do}\;{}\<[25]%
\>[25]{}\Varid{y}\leftarrow \Varid{power'}\;(\mathbin{-}\Varid{n})\;\Varid{x}{}\<[E]%
\\
\>[25]{}\Varid{return}\;(\mathrm{1}\mathbin{/}\Varid{y})){}\<[E]%
\\
\>[B]{}\hsindent{3}{}\<[3]%
\>[3]{}\mathbf{else}\;\mathbf{if}\;\Varid{n}\longeq \mathrm{0}\;\mathbf{then}{}\<[E]%
\\
\>[3]{}\hsindent{2}{}\<[5]%
\>[5]{}\Varid{return}\;\mathrm{1}{}\<[E]%
\\
\>[B]{}\hsindent{3}{}\<[3]%
\>[3]{}\mathbf{else}\;\mathbf{if}\;\Varid{even}\;\Varid{n}\;\mathbf{then}{}\<[E]%
\\
\>[3]{}\hsindent{2}{}\<[5]%
\>[5]{}\mathbf{do}\;{}\<[9]%
\>[9]{}\Varid{y}\leftarrow \Varid{power'}\;(\Varid{n}\rmdiv\mathrm{2})\;\Varid{x}{}\<[E]%
\\
\>[9]{}\Varid{return}\;(\Varid{y}\times\Varid{y}){}\<[E]%
\\
\>[B]{}\hsindent{3}{}\<[3]%
\>[3]{}\mathbf{else}{}\<[E]%
\\
\>[3]{}\hsindent{2}{}\<[5]%
\>[5]{}\mathbf{do}\;{}\<[9]%
\>[9]{}\Varid{y}\leftarrow \Varid{power'}\;(\Varid{n}\mathbin{-}\mathrm{1})\;\Varid{x}{}\<[E]%
\\
\>[9]{}\Varid{return}\;(\Varid{x}\times\Varid{y}){}\<[E]%
\\[\blanklineskip]%
\>[B]{}\Varid{power''}\mathbin{::}{}\<[13]%
\>[13]{}\Conid{Int}\to \Conid{Dp}\;\Conid{Float}\to \Conid{Dp}\;\Conid{Float}{}\<[E]%
\\
\>[B]{}\Varid{power''}\;\Varid{n}\;\Varid{x}{}\<[14]%
\>[14]{}\mathrel{=}{}\<[14E]%
\>[17]{}\Varid{option}\;\mathrm{0}\;(\lambda \Varid{y}\to \Varid{y})\;(\Varid{power'}\;\Varid{n}\;\Varid{x}){}\<[E]%
\ColumnHook
\end{hscode}\resethooks
The term of type \ensuremath{\Conid{Dp}\;\Conid{Float}} generated by
evaluating \ensuremath{\Varid{power}\;(\mathbin{-}\mathrm{6})\;\Varid{x}} is large and unscrutable:
\begin{hscode}\SaveRestoreHook
\column{B}{@{}>{\hspre}l<{\hspost}@{}}%
\column{11}{@{}>{\hspre}l<{\hspost}@{}}%
\column{16}{@{}>{\hspre}l<{\hspost}@{}}%
\column{33}{@{}>{\hspre}l<{\hspost}@{}}%
\column{53}{@{}>{\hspre}l<{\hspost}@{}}%
\column{54}{@{}>{\hspre}l<{\hspost}@{}}%
\column{E}{@{}>{\hspre}l<{\hspost}@{}}%
\>[B]{}(((\Varid{fst}\;((\Varid{x}\longeq \mathrm{0.0})\mathbin{?}(((\Conid{False}\mathbin{?}(\Conid{True},\Conid{False})),(\Conid{False}\mathbin{?}{}\<[E]%
\\
\>[B]{}(\Varid{undef},\Varid{undef}))),(\Conid{True},(\mathrm{1.0}\mathbin{/}((\Varid{x}\times((\Varid{x}\times\mathrm{1.0})\times(\Varid{x}\times{}\<[E]%
\\
\>[B]{}\mathrm{1.0})))\times(\Varid{x}\times((\Varid{x}\times\mathrm{1.0})\times(\Varid{x}\times\mathrm{1.0})))))))))\mathbin{?}(\Conid{True},{}\<[E]%
\\
\>[B]{}\Conid{False}))\mathbin{?}(((\Varid{fst}\;((\Varid{x}\longeq \mathrm{0.0})\mathbin{?}{}\<[33]%
\>[33]{}(((\Conid{False}\mathbin{?}(\Conid{True},\Conid{False})),{}\<[E]%
\\
\>[B]{}(\Conid{False}\mathbin{?}{}\<[11]%
\>[11]{}(\Varid{undef},\Varid{undef}))),(\Conid{True},(\mathrm{1.0}\mathbin{/}((\Varid{x}\times((\Varid{x}\times\mathrm{1.0})\times{}\<[E]%
\\
\>[B]{}(\Varid{x}\times\mathrm{1.0})))\times(\Varid{x}\times((\Varid{x}\times\mathrm{1.0})\times(\Varid{x}\times\mathrm{1.0})))))))))\mathbin{?}{}\<[54]%
\>[54]{}((\Varid{snd}{}\<[E]%
\\
\>[B]{}((\Varid{x}\longeq \mathrm{0.0})\mathbin{?}{}\<[16]%
\>[16]{}(((\Conid{False}\mathbin{?}(\Conid{True},\Conid{False})),(\Conid{False}\mathbin{?}{}\<[53]%
\>[53]{}(\Varid{undef},{}\<[E]%
\\
\>[B]{}\Varid{undef}))),(\Conid{True},(\mathrm{1.0}\mathbin{/}((\Varid{x}\times((\Varid{x}\times\mathrm{1.0})\times(\Varid{x}\times\mathrm{1.0})))\times{}\<[E]%
\\
\>[B]{}(\Varid{x}\times((\Varid{x}\times\mathrm{1.0})\times(\Varid{x}\times\mathrm{1.0}))))))))),\Varid{undef})),\mathrm{0.0})){}\<[E]%
\ColumnHook
\end{hscode}\resethooks
Before, evaluating \ensuremath{\Varid{power}} yielded a term essentially in normal
form.  However, here rewrite rules need to be repeatedly applied,
as described in Section~\ref{sec:implementation}.
After applying these rules, common subexpression
elimination yields the same structure as in the previous subsection,
from which the same C code is generated.

Here we have described normalisation via rewriting, but some EDSLs
achieve normalisation via smart constructors,
which ensure deep terms are always in normal form
\citep{rompf2012lightweight}; the two techniques are roughly equivalent.


Hence, an advantage of the EDSL approach---that it generates terms
essentially in normal form---turns out to apply sometimes but not
others. It appears to often work for functions and products, but to
fail for sums.  In such situations, separate normalisation is
required. This is one reason why we do not consider normalisation as
required by QDSL to be particularly onerous.

Here are points of comparison between the two approaches.
\begin{itemize}

\item Both QDSL and EDSL can exploit notational conveniences in the
host language. The example here exploits Haskell \ensuremath{\mathbf{do}} notation; the
embedding of SQL in F\# by \citet{cheney:linq} expoits F\# sequence
notation. For EDSL, exploiting \ensuremath{\mathbf{do}} notation just requires
instantiating \ensuremath{\Varid{return}} and \ensuremath{(\bind )} correctly. For QDSL, it is
also necessary for the translator to recognise and expand
\ensuremath{\mathbf{do}} notation and to substitute appropriate instances of
\ensuremath{\Varid{return}} and \ensuremath{(\bind )}.

\item As this example shows, sometimes both QDSL and EDSL may require
normalisation.  As mentioned previously, for QDSLs the cost of
building a normaliser might be amortised across several applications.
In constrast, each EDSL usually has a distinct deep representation and
so requires a distinct normaliser.

\end{itemize}

\subsection{The deep embedding}
\label{subsec:deep}

Recall that a value of type \ensuremath{\Conid{Dp}\;\Varid{a}} represents a term of type \ensuremath{\Varid{a}},
and is called a deep embedding.
\begin{hscode}\SaveRestoreHook
\column{B}{@{}>{\hspre}l<{\hspost}@{}}%
\column{3}{@{}>{\hspre}l<{\hspost}@{}}%
\column{13}{@{}>{\hspre}c<{\hspost}@{}}%
\column{13E}{@{}l@{}}%
\column{17}{@{}>{\hspre}l<{\hspost}@{}}%
\column{E}{@{}>{\hspre}l<{\hspost}@{}}%
\>[B]{}\mathbf{data}\;\Conid{Dp}\;\Varid{a}\;\mathbf{where}{}\<[E]%
\\
\>[B]{}\hsindent{3}{}\<[3]%
\>[3]{}\Conid{LitB}{}\<[13]%
\>[13]{}\mathbin{::}{}\<[13E]%
\>[17]{}\Conid{Bool}\to \Conid{Dp}\;\Conid{Bool}{}\<[E]%
\\
\>[B]{}\hsindent{3}{}\<[3]%
\>[3]{}\Conid{LitI}{}\<[13]%
\>[13]{}\mathbin{::}{}\<[13E]%
\>[17]{}\Conid{Int}\to \Conid{Dp}\;\Conid{Int}{}\<[E]%
\\
\>[B]{}\hsindent{3}{}\<[3]%
\>[3]{}\Conid{LitF}{}\<[13]%
\>[13]{}\mathbin{::}{}\<[13E]%
\>[17]{}\Conid{Float}\to \Conid{Dp}\;\Conid{Float}{}\<[E]%
\\
\>[B]{}\hsindent{3}{}\<[3]%
\>[3]{}\Conid{If}{}\<[13]%
\>[13]{}\mathbin{::}{}\<[13E]%
\>[17]{}\Conid{Dp}\;\Conid{Bool}\to \Conid{Dp}\;\Varid{a}\to \Conid{Dp}\;\Varid{a}\to \Conid{Dp}\;\Varid{a}{}\<[E]%
\\
\>[B]{}\hsindent{3}{}\<[3]%
\>[3]{}\Conid{While}{}\<[13]%
\>[13]{}\mathbin{::}{}\<[13E]%
\>[17]{}(\Conid{Dp}\;\Varid{a}\to \Conid{Dp}\;\Conid{Bool})\to {}\<[E]%
\\
\>[17]{}(\Conid{Dp}\;\Varid{a}\to \Conid{Dp}\;\Varid{a})\to \Conid{Dp}\;\Varid{a}\to \Conid{Dp}\;\Varid{a}{}\<[E]%
\\
\>[B]{}\hsindent{3}{}\<[3]%
\>[3]{}\Conid{Pair}{}\<[13]%
\>[13]{}\mathbin{::}{}\<[13E]%
\>[17]{}\Conid{Dp}\;\Varid{a}\to \Conid{Dp}\;\Varid{b}\to \Conid{Dp}\;(\Varid{a},\Varid{b}){}\<[E]%
\\
\>[B]{}\hsindent{3}{}\<[3]%
\>[3]{}\Conid{Fst}{}\<[13]%
\>[13]{}\mathbin{::}{}\<[13E]%
\>[17]{}\Conid{Rep}\;\Varid{b}\Rightarrow \Conid{Dp}\;(\Varid{a},\Varid{b})\to \Conid{Dp}\;\Varid{a}{}\<[E]%
\\
\>[B]{}\hsindent{3}{}\<[3]%
\>[3]{}\Conid{Snd}{}\<[13]%
\>[13]{}\mathbin{::}{}\<[13E]%
\>[17]{}\Conid{Rep}\;\Varid{a}\Rightarrow \Conid{Dp}\;(\Varid{a},\Varid{b})\to \Conid{Dp}\;\Varid{b}{}\<[E]%
\\
\>[B]{}\hsindent{3}{}\<[3]%
\>[3]{}\Conid{Prim1}{}\<[13]%
\>[13]{}\mathbin{::}{}\<[13E]%
\>[17]{}\Conid{Rep}\;\Varid{a}\Rightarrow \Conid{String}\to \Conid{Dp}\;\Varid{a}\to \Conid{Dp}\;\Varid{b}{}\<[E]%
\\
\>[B]{}\hsindent{3}{}\<[3]%
\>[3]{}\Conid{Prim2}{}\<[13]%
\>[13]{}\mathbin{::}{}\<[13E]%
\>[17]{}(\Conid{Rep}\;\Varid{a},\Conid{Rep}\;\Varid{b})\Rightarrow {}\<[E]%
\\
\>[17]{}\Conid{String}\to \Conid{Dp}\;\Varid{a}\to \Conid{Dp}\;\Varid{b}\to \Conid{Dp}\;\Varid{c}{}\<[E]%
\\
\>[B]{}\hsindent{3}{}\<[3]%
\>[3]{}\Conid{MkArr}{}\<[13]%
\>[13]{}\mathbin{::}{}\<[13E]%
\>[17]{}\Conid{Dp}\;\Conid{Int}\to (\Conid{Dp}\;\Conid{Int}\to \Conid{Dp}\;\Varid{a})\to \Conid{Dp}\;(\Conid{Arr}\;\Varid{a}){}\<[E]%
\\
\>[B]{}\hsindent{3}{}\<[3]%
\>[3]{}\Conid{LnArr}{}\<[13]%
\>[13]{}\mathbin{::}{}\<[13E]%
\>[17]{}\Conid{Rep}\;\Varid{a}\Rightarrow \Conid{Dp}\;(\Conid{Arr}\;\Varid{a})\to \Conid{Dp}\;\Conid{Int}{}\<[E]%
\\
\>[B]{}\hsindent{3}{}\<[3]%
\>[3]{}\Conid{IxArr}{}\<[13]%
\>[13]{}\mathbin{::}{}\<[13E]%
\>[17]{}\Conid{Dp}\;(\Conid{Arr}\;\Varid{a})\to \Conid{Dp}\;\Conid{Int}\to \Conid{Dp}\;\Varid{a}{}\<[E]%
\\
\>[B]{}\hsindent{3}{}\<[3]%
\>[3]{}\Conid{Save}{}\<[13]%
\>[13]{}\mathbin{::}{}\<[13E]%
\>[17]{}\Conid{Dp}\;\Varid{a}\to \Conid{Dp}\;\Varid{a}{}\<[E]%
\\
\>[B]{}\hsindent{3}{}\<[3]%
\>[3]{}\Conid{Let}{}\<[13]%
\>[13]{}\mathbin{::}{}\<[13E]%
\>[17]{}\Conid{Rep}\;\Varid{a}\Rightarrow \Conid{Dp}\;\Varid{a}\to (\Conid{Dp}\;\Varid{a}\to \Conid{Dp}\;\Varid{b})\to \Conid{Dp}\;\Varid{b}{}\<[E]%
\\
\>[B]{}\hsindent{3}{}\<[3]%
\>[3]{}\Conid{Variable}{}\<[13]%
\>[13]{}\mathbin{::}{}\<[13E]%
\>[17]{}\Conid{String}\to \Conid{Dp}\;\Varid{a}{}\<[E]%
\ColumnHook
\end{hscode}\resethooks
Type \ensuremath{\Conid{Dp}} represents a low level, pure functional language
with a straightforward translation to C. It uses higher-order
abstract syntax (HOAS) to represent constructs with variable binding
\citet{hoas}.
Our code obeys the invariant that we only write \ensuremath{\Conid{Dp}\;\Varid{a}}
when \ensuremath{\Conid{Rep}\;\Varid{a}} holds, that is, when type \ensuremath{\Varid{a}} is representable.

The deep embedding has boolean, integer, and floating point literals,
conditionals, while loops, pairs, primitives, arrays,
and special-purpose constructs to disable normalisation,
for let binding, and for variables.
Constructs \ensuremath{\Conid{LitB}}, \ensuremath{\Conid{LitI}}, \ensuremath{\Conid{LitF}} build literals;
\ensuremath{\Conid{If}} builds a conditional.
\ensuremath{\Conid{While}} corresponds to \ensuremath{\Varid{while}} in Section~\ref{subsec:while};
\ensuremath{\Conid{Pair}}, \ensuremath{\Conid{Fst}}, and \ensuremath{\Conid{Snd}} build and decompose pairs;
\ensuremath{\Conid{Prim1}} and \ensuremath{\Conid{Prim2}} represent primitive
operations, where the string is the name of the operation;
\ensuremath{\Conid{MkArr}}, \ensuremath{\Conid{LnArr}}, and \ensuremath{\Conid{IxArr}}
correspond to the array operations in Section~\ref{subsec:arrays};
\ensuremath{\Conid{Save}} corresponds to \ensuremath{\Varid{save}} in Section~\ref{subsec:subformula};
\ensuremath{\Conid{Let}} corresponds to let binding,
and \ensuremath{\Conid{Variable}} is used when translating HOAS to C code.

\subsection{Class \ensuremath{\Conid{Syn}}}
\label{subsec:syn}

We introduce a type class \ensuremath{\Conid{Syn}} that allows us to convert
shallow embeddings to and from deep embeddings.
\begin{hscode}\SaveRestoreHook
\column{B}{@{}>{\hspre}l<{\hspost}@{}}%
\column{3}{@{}>{\hspre}l<{\hspost}@{}}%
\column{11}{@{}>{\hspre}c<{\hspost}@{}}%
\column{11E}{@{}l@{}}%
\column{15}{@{}>{\hspre}l<{\hspost}@{}}%
\column{E}{@{}>{\hspre}l<{\hspost}@{}}%
\>[B]{}\mathbf{class}\;\Conid{Rep}\;(\Conid{Internal}\;\Varid{a})\Rightarrow \Conid{Syn}\;\Varid{a}\;\mathbf{where}{}\<[E]%
\\
\>[B]{}\hsindent{3}{}\<[3]%
\>[3]{}\mathbf{type}\;\Conid{Internal}\;\Varid{a}{}\<[E]%
\\
\>[B]{}\hsindent{3}{}\<[3]%
\>[3]{}\Varid{toDp}{}\<[11]%
\>[11]{}\mathbin{::}{}\<[11E]%
\>[15]{}\Varid{a}\to \Conid{Dp}\;(\Conid{Internal}\;\Varid{a}){}\<[E]%
\\
\>[B]{}\hsindent{3}{}\<[3]%
\>[3]{}\Varid{fromDp}{}\<[11]%
\>[11]{}\mathbin{::}{}\<[11E]%
\>[15]{}\Conid{Dp}\;(\Conid{Internal}\;\Varid{a})\to \Varid{a}{}\<[E]%
\ColumnHook
\end{hscode}\resethooks
Type \ensuremath{\Conid{Internal}} is a GHC type family \citep{type-families}.  Functions
\ensuremath{\Varid{toDp}} and \ensuremath{\Varid{fromDp}} translate between the shallow embedding \ensuremath{\Varid{a}} and the
deep embedding \ensuremath{\Conid{Dp}\;(\Conid{Internal}\;\Varid{a})}.

The first instance of \ensuremath{\Conid{Syn}} is \ensuremath{\Conid{Dp}} itself, and is straightforward.
\begin{hscode}\SaveRestoreHook
\column{B}{@{}>{\hspre}l<{\hspost}@{}}%
\column{3}{@{}>{\hspre}l<{\hspost}@{}}%
\column{11}{@{}>{\hspre}c<{\hspost}@{}}%
\column{11E}{@{}l@{}}%
\column{14}{@{}>{\hspre}l<{\hspost}@{}}%
\column{E}{@{}>{\hspre}l<{\hspost}@{}}%
\>[B]{}\mathbf{instance}\;\Conid{Rep}\;\Varid{a}\Rightarrow \Conid{Syn}\;(\Conid{Dp}\;\Varid{a})\;\mathbf{where}{}\<[E]%
\\
\>[B]{}\hsindent{3}{}\<[3]%
\>[3]{}\mathbf{type}\;\Conid{Internal}\;(\Conid{Dp}\;\Varid{a})\mathrel{=}\Varid{a}{}\<[E]%
\\
\>[B]{}\hsindent{3}{}\<[3]%
\>[3]{}\Varid{toDp}{}\<[11]%
\>[11]{}\mathrel{=}{}\<[11E]%
\>[14]{}\Varid{id}{}\<[E]%
\\
\>[B]{}\hsindent{3}{}\<[3]%
\>[3]{}\Varid{fromDp}{}\<[11]%
\>[11]{}\mathrel{=}{}\<[11E]%
\>[14]{}\Varid{id}{}\<[E]%
\ColumnHook
\end{hscode}\resethooks
Our representation of a run-time \ensuremath{\Conid{Bool}} will have type \ensuremath{\Conid{Dp}\;\Conid{Bool}} in
both the deep and shallow embeddings, and similarly for \ensuremath{\Conid{Int}} and
\ensuremath{\Conid{Float}}.

We do not code the target language using its constructs
directly. Instead, for each constructor we define a corresponding
``smart constructor'' using class \ensuremath{\Conid{Syn}}.
\begin{hscode}\SaveRestoreHook
\column{B}{@{}>{\hspre}l<{\hspost}@{}}%
\column{30}{@{}>{\hspre}l<{\hspost}@{}}%
\column{E}{@{}>{\hspre}l<{\hspost}@{}}%
\>[B]{}\Varid{true},\Varid{false}\mathbin{::}\Conid{Dp}\;\Conid{Bool}{}\<[E]%
\\
\>[B]{}\Varid{true}\mathrel{=}\Conid{LitB}\;\Conid{True}{}\<[E]%
\\
\>[B]{}\Varid{false}\mathrel{=}\Conid{LitB}\;\Conid{False}{}\<[E]%
\\[\blanklineskip]%
\>[B]{}(\mathbin{?})\mathbin{::}\Conid{Syn}\;\Varid{a}\Rightarrow \Conid{Dp}\;\Conid{Bool}\to (\Varid{a},\Varid{a})\to \Varid{a}{}\<[E]%
\\
\>[B]{}\Varid{c}\mathbin{?}(\Varid{t},\Varid{e})\mathrel{=}\Varid{fromDp}\;(\Conid{If}\;\Varid{c}\;(\Varid{toDp}\;\Varid{t})\;(\Varid{toDp}\;\Varid{e})){}\<[E]%
\\[\blanklineskip]%
\>[B]{}\Varid{while}\mathbin{::}\Conid{Syn}\;\Varid{a}\Rightarrow (\Varid{a}\to \Conid{Dp}\;\Conid{Bool})\to (\Varid{a}\to \Varid{a})\to \Varid{a}\to \Varid{a}{}\<[E]%
\\
\>[B]{}\Varid{while}\;\Varid{c}\;\Varid{b}\;\Varid{i}\mathrel{=}\Varid{fromDp}\;(\Conid{While}\;{}\<[30]%
\>[30]{}(\Varid{c}\hsdot{\circ }{.}\Varid{fromDp})\;{}\<[E]%
\\
\>[30]{}(\Varid{toDp}\hsdot{\circ }{.}\Varid{b}\hsdot{\circ }{.}\Varid{fromDp})\;{}\<[E]%
\\
\>[30]{}(\Varid{toDp}\;\Varid{i})){}\<[E]%
\ColumnHook
\end{hscode}\resethooks

Numbers are made convenient to manipulate via overloading.
\begin{hscode}\SaveRestoreHook
\column{B}{@{}>{\hspre}l<{\hspost}@{}}%
\column{3}{@{}>{\hspre}l<{\hspost}@{}}%
\column{10}{@{}>{\hspre}c<{\hspost}@{}}%
\column{10E}{@{}l@{}}%
\column{13}{@{}>{\hspre}l<{\hspost}@{}}%
\column{E}{@{}>{\hspre}l<{\hspost}@{}}%
\>[B]{}\mathbf{instance}\;\Conid{Num}\;(\Conid{Dp}\;\Conid{Int})\;\mathbf{where}{}\<[E]%
\\
\>[B]{}\hsindent{3}{}\<[3]%
\>[3]{}\Varid{a}\mathbin{+}\Varid{b}{}\<[10]%
\>[10]{}\mathrel{=}{}\<[10E]%
\>[13]{}\Conid{Prim2}\;\text{\tt \char34 (+)\char34}\;\Varid{a}\;\Varid{b}{}\<[E]%
\\
\>[B]{}\hsindent{3}{}\<[3]%
\>[3]{}\Varid{a}\mathbin{-}\Varid{b}{}\<[10]%
\>[10]{}\mathrel{=}{}\<[10E]%
\>[13]{}\Conid{Prim2}\;\text{\tt \char34 (-)\char34}\;\Varid{a}\;\Varid{b}{}\<[E]%
\\
\>[B]{}\hsindent{3}{}\<[3]%
\>[3]{}\Varid{a}\times\Varid{b}{}\<[10]%
\>[10]{}\mathrel{=}{}\<[10E]%
\>[13]{}\Conid{Prim2}\;\text{\tt \char34 (*)\char34}\;\Varid{a}\;\Varid{b}{}\<[E]%
\\
\>[B]{}\hsindent{3}{}\<[3]%
\>[3]{}\Varid{fromInteger}\;\Varid{a}\mathrel{=}\Conid{LitI}\;(\Varid{fromInteger}\;\Varid{a}){}\<[E]%
\ColumnHook
\end{hscode}\resethooks
With this declaration, \ensuremath{\mathrm{1}\mathbin{+}\mathrm{2}\mathbin{::}\Conid{Dp}\;\Conid{Int}} evaluates to
\[
\ensuremath{\Conid{Prim2}\;\text{\tt \char34 (+)\char34}\;(\Conid{LitI}\;\mathrm{1})\;(\Conid{LitI}\;\mathrm{2})},
\]
permitting code executed at generation-time and run-time to
appear identical.  A similar declaration works for \ensuremath{\Conid{Float}}.

Comparison also benefits from smart constructors.
\begin{hscode}\SaveRestoreHook
\column{B}{@{}>{\hspre}l<{\hspost}@{}}%
\column{E}{@{}>{\hspre}l<{\hspost}@{}}%
\>[B]{}(\mathbin{{.}{\longeq }{.}})\mathbin{::}(\Conid{Syn}\;\Varid{a},\Conid{Eq}\;(\Conid{Internal}\;\Varid{a}))\Rightarrow \Varid{a}\to \Varid{a}\to \Conid{Dp}\;\Conid{Bool}{}\<[E]%
\\
\>[B]{}\Varid{a}\mathbin{{.}{\longeq }{.}}\Varid{b}\mathrel{=}\Conid{Prim2}\;\text{\tt \char34 (==)\char34}\;(\Varid{toDp}\;\Varid{a})\;(\Varid{toDp}\;\Varid{b}){}\<[E]%
\\[\blanklineskip]%
\>[B]{}(\mathbin{{.}{\mathbin{<}}{.}})\mathbin{::}(\Conid{Syn}\;\Varid{a},\Conid{Ord}\;(\Conid{Internal}\;\Varid{a}))\Rightarrow \Varid{a}\to \Varid{a}\to \Conid{Dp}\;\Conid{Bool}{}\<[E]%
\\
\>[B]{}\Varid{a}\mathbin{{.}{\mathbin{<}}{.}}\Varid{b}\mathrel{=}\Conid{Prim2}\;\text{\tt \char34 (<)\char34}\;(\Varid{toDp}\;\Varid{a})\;(\Varid{toDp}\;\Varid{b}){}\<[E]%
\ColumnHook
\end{hscode}\resethooks
Overloading cannot apply here, because Haskell requires
\ensuremath{(\longeq )} return a result of type \ensuremath{\Conid{Bool}}, while \ensuremath{(\mathbin{{.}{\longeq }{.}})} returns
a result of type \ensuremath{\Conid{Dp}\;\Conid{Bool}}, and similarly for \ensuremath{(\mathbin{{.}{\mathbin{<}}{.}})}.

Here is how to compute the minimum of two values.
\begin{hscode}\SaveRestoreHook
\column{B}{@{}>{\hspre}l<{\hspost}@{}}%
\column{15}{@{}>{\hspre}c<{\hspost}@{}}%
\column{15E}{@{}l@{}}%
\column{19}{@{}>{\hspre}l<{\hspost}@{}}%
\column{E}{@{}>{\hspre}l<{\hspost}@{}}%
\>[B]{}\Varid{minim}{}\<[15]%
\>[15]{}\mathbin{::}{}\<[15E]%
\>[19]{}(\Conid{Syn}\;\Varid{a},\Conid{Ord}\;(\Conid{Internal}\;\Varid{a}))\Rightarrow \Varid{a}\to \Varid{a}\to \Varid{a}{}\<[E]%
\\
\>[B]{}\Varid{minim}\;\Varid{x}\;\Varid{y}{}\<[15]%
\>[15]{}\mathrel{=}{}\<[15E]%
\>[19]{}(\Varid{x}\mathbin{{.}{\mathbin{<}}{.}}\Varid{y})\mathbin{?}(\Varid{x},\Varid{y}){}\<[E]%
\ColumnHook
\end{hscode}\resethooks

\subsection{Embedding pairs}

Host language pairs in the shallow embedding
correspond to target language pairs in the deep embedding.
\begin{hscode}\SaveRestoreHook
\column{B}{@{}>{\hspre}l<{\hspost}@{}}%
\column{3}{@{}>{\hspre}l<{\hspost}@{}}%
\column{9}{@{}>{\hspre}l<{\hspost}@{}}%
\column{15}{@{}>{\hspre}c<{\hspost}@{}}%
\column{15E}{@{}l@{}}%
\column{18}{@{}>{\hspre}l<{\hspost}@{}}%
\column{E}{@{}>{\hspre}l<{\hspost}@{}}%
\>[B]{}\mathbf{instance}\;(\Conid{Syn}\;\Varid{a},\Conid{Syn}\;\Varid{b})\Rightarrow \Conid{Syn}\;(\Varid{a},\Varid{b})\;\mathbf{where}{}\<[E]%
\\
\>[B]{}\hsindent{3}{}\<[3]%
\>[3]{}\mathbf{type}\;{}\<[9]%
\>[9]{}\Conid{Internal}\;(\Varid{a},\Varid{b})\mathrel{=}(\Conid{Internal}\;\Varid{a},\Conid{Internal}\;\Varid{b}){}\<[E]%
\\
\>[B]{}\hsindent{3}{}\<[3]%
\>[3]{}\Varid{toDp}\;(\Varid{a},\Varid{b}){}\<[15]%
\>[15]{}\mathrel{=}{}\<[15E]%
\>[18]{}\Conid{Pair}\;(\Varid{toDp}\;\Varid{a})\;(\Varid{toDp}\;\Varid{b}){}\<[E]%
\\
\>[B]{}\hsindent{3}{}\<[3]%
\>[3]{}\Varid{fromDp}\;\Varid{p}{}\<[15]%
\>[15]{}\mathrel{=}{}\<[15E]%
\>[18]{}(\Varid{fromDp}\;(\Conid{Fst}\;\Varid{p}),\Varid{fromDp}\;(\Conid{Snd}\;\Varid{p})){}\<[E]%
\ColumnHook
\end{hscode}\resethooks
This permits us to manipulate pairs as normal, with \ensuremath{(\Varid{a},\Varid{b})}, \ensuremath{\Varid{fst}\;\Varid{a}},
and \ensuremath{\Varid{snd}\;\Varid{a}}.  Argument \ensuremath{\Varid{p}} is duplicated in the definition of
\ensuremath{\Varid{fromDp}}, which may require common subexpression elimination
as discussed in Section~\ref{subsec:e-power}.

We have now developed sufficient machinery to define a \ensuremath{\Varid{for}} loop
in terms of a \ensuremath{\Varid{while}} loop.
\begin{hscode}\SaveRestoreHook
\column{B}{@{}>{\hspre}l<{\hspost}@{}}%
\column{27}{@{}>{\hspre}l<{\hspost}@{}}%
\column{E}{@{}>{\hspre}l<{\hspost}@{}}%
\>[B]{}\Varid{for}\mathbin{::}\Conid{Syn}\;\Varid{a}\Rightarrow \Conid{Dp}\;\Conid{Int}\to \Varid{a}\to (\Conid{Dp}\;\Conid{Int}\to \Varid{a}\to \Varid{a})\to \Varid{a}{}\<[E]%
\\
\>[B]{}\Varid{for}\;\Varid{n}\;\Varid{s}_{\mathrm{0}}\;\Varid{b}\mathrel{=}\Varid{snd}\;(\Varid{while}\;{}\<[27]%
\>[27]{}(\lambda (\Varid{i},\Varid{s})\to \Varid{i}\mathbin{{.}{\mathbin{<}}{.}}\Varid{n})\;{}\<[E]%
\\
\>[27]{}(\lambda (\Varid{i},\Varid{s})\to (\Varid{i}\mathbin{+}\mathrm{1},\Varid{b}\;\Varid{i}\;\Varid{s}))\;{}\<[E]%
\\
\>[27]{}(\mathrm{0},\Varid{s}_{\mathrm{0}})){}\<[E]%
\ColumnHook
\end{hscode}\resethooks
The state of the \ensuremath{\Varid{while}} loop is a pair consisting of a counter and
the state of the \ensuremath{\Varid{for}} loop. The body \ensuremath{\Varid{b}} of the \ensuremath{\Varid{for}} loop is a function
that expects both the counter and the state of the \ensuremath{\Varid{for}} loop.
The counter is discarded when the loop is complete, and the final state
of the \ensuremath{\Varid{for}} loop returned.

Thanks to our machinery, the above definition uses only ordinary Haskell
pairs. The condition and body of the \ensuremath{\Varid{while}} loop pattern match on the
state using ordinary pair syntax, and the initial state is constructed
as an ordinary pair.

\subsection{Embedding undefined}
\label{subsec:undef}

For the next section, which defines an analogue of the \ensuremath{\Conid{Maybe}} type, it
will prove convenient to work with types which have a distinguished
value at each type, which we call \ensuremath{\Varid{undef}}.


It is straightforward to define a type class \ensuremath{\Conid{Undef}}, where type \ensuremath{\Varid{a}}
belongs to \ensuremath{\Conid{Undef}} if it belongs to \ensuremath{\Conid{Syn}} and it has an
undefined value.
\begin{hscode}\SaveRestoreHook
\column{B}{@{}>{\hspre}l<{\hspost}@{}}%
\column{3}{@{}>{\hspre}l<{\hspost}@{}}%
\column{E}{@{}>{\hspre}l<{\hspost}@{}}%
\>[B]{}\mathbf{class}\;\Conid{Syn}\;\Varid{a}\Rightarrow \Conid{Undef}\;\Varid{a}\;\mathbf{where}{}\<[E]%
\\
\>[B]{}\hsindent{3}{}\<[3]%
\>[3]{}\Varid{undef}\mathbin{::}\Varid{a}{}\<[E]%
\\[\blanklineskip]%
\>[B]{}\mathbf{instance}\;\Conid{Undef}\;(\Conid{Dp}\;\Conid{Bool})\;\mathbf{where}{}\<[E]%
\\
\>[B]{}\hsindent{3}{}\<[3]%
\>[3]{}\Varid{undef}\mathrel{=}\Varid{false}{}\<[E]%
\\[\blanklineskip]%
\>[B]{}\mathbf{instance}\;\Conid{Undef}\;(\Conid{Dp}\;\Conid{Int})\;\mathbf{where}{}\<[E]%
\\
\>[B]{}\hsindent{3}{}\<[3]%
\>[3]{}\Varid{undef}\mathrel{=}\mathrm{0}{}\<[E]%
\\[\blanklineskip]%
\>[B]{}\mathbf{instance}\;\Conid{Undef}\;(\Conid{Dp}\;\Conid{Float})\;\mathbf{where}{}\<[E]%
\\
\>[B]{}\hsindent{3}{}\<[3]%
\>[3]{}\Varid{undef}\mathrel{=}\mathrm{0}{}\<[E]%
\\[\blanklineskip]%
\>[B]{}\mathbf{instance}\;(\Conid{Undef}\;\Varid{a},\Conid{Undef}\;\Varid{b})\Rightarrow \Conid{Undef}\;(\Varid{a},\Varid{b})\;\mathbf{where}{}\<[E]%
\\
\>[B]{}\hsindent{3}{}\<[3]%
\>[3]{}\Varid{undef}\mathrel{=}(\Varid{undef},\Varid{undef}){}\<[E]%
\ColumnHook
\end{hscode}\resethooks

For example,
\begin{hscode}\SaveRestoreHook
\column{B}{@{}>{\hspre}l<{\hspost}@{}}%
\column{9}{@{}>{\hspre}c<{\hspost}@{}}%
\column{9E}{@{}l@{}}%
\column{13}{@{}>{\hspre}l<{\hspost}@{}}%
\column{E}{@{}>{\hspre}l<{\hspost}@{}}%
\>[B]{}(\mathbin{/\#}){}\<[9]%
\>[9]{}\mathbin{::}{}\<[9E]%
\>[13]{}\Conid{Dp}\;\Conid{Float}\to \Conid{Dp}\;\Conid{Float}\to \Conid{Dp}\;\Conid{Float}{}\<[E]%
\\
\>[B]{}\Varid{x}\mathbin{/\#}\Varid{y}{}\<[9]%
\>[9]{}\mathrel{=}{}\<[9E]%
\>[13]{}(\Varid{y}\mathbin{{.}{\longeq }{.}}\mathrm{0})\mathbin{?}(\Varid{undef},\Varid{x}\mathbin{/}\Varid{y}){}\<[E]%
\ColumnHook
\end{hscode}\resethooks
behaves as division, save that when the divisor is zero
it returns the undefined value of type \ensuremath{\Conid{Float}}, which
is also zero.

\citet{svenningsson:combining} claim that it is not possible to support
\ensuremath{\Varid{undef}} without changing the deep embedding, but here we have defined \ensuremath{\Varid{undef}}
entirely as a shallow embedding.  (It appears they underestimated the
power of their own technique!)

\subsection{Embedding option}
\label{subsec:opt}

We now explain in detail the \ensuremath{\Conid{Opt}} type seen in Section~\ref{subsec:maybe}.

The deep-and-shallow technique represents deep embeddding
\ensuremath{\Conid{Dp}\;(\Varid{a},\Varid{b})} by shallow embedding \ensuremath{(\Conid{Dp}\;\Varid{a},\Conid{Dp}\;\Varid{b})}.  Hence, it is tempting to
represent \ensuremath{\Conid{Dp}\;(\Conid{Maybe}\;\Varid{a})} by \ensuremath{\Conid{Maybe}\;(\Conid{Dp}\;\Varid{a})}, but this cannot work,
because \ensuremath{\Varid{fromDp}} would have to decide at generation-time whether to
return \ensuremath{\Conid{Just}} or \ensuremath{\Conid{Nothing}}, but which to use is not known until
run-time.

Instead, \citet{svenningsson:combining} represent values of type
\ensuremath{\Conid{Maybe}\;\Varid{a}} by the type \ensuremath{\Conid{Opt'}\;\Varid{a}}, which pairs a boolean with a value of
type \ensuremath{\Varid{a}}.  For a value corresponding to \ensuremath{\Conid{Just}\;\Varid{x}}, the boolean is true
and the value is \ensuremath{\Varid{x}}, while for one corresponding to \ensuremath{\Conid{Nothing}}, the
boolean is false and the value is \ensuremath{\Varid{undef}}.  We define \ensuremath{\Varid{some'}},
\ensuremath{\Varid{none'}}, and \ensuremath{\Varid{option'}} as the analogues of \ensuremath{\Conid{Just}}, \ensuremath{\Conid{Nothing}}, and
\ensuremath{\Varid{maybe}}.  The \ensuremath{\Conid{Syn}} instance is straightforward, mapping options to
and from the pairs already defined for \ensuremath{\Conid{Dp}}.
\begin{hscode}\SaveRestoreHook
\column{B}{@{}>{\hspre}l<{\hspost}@{}}%
\column{3}{@{}>{\hspre}l<{\hspost}@{}}%
\column{16}{@{}>{\hspre}c<{\hspost}@{}}%
\column{16E}{@{}l@{}}%
\column{20}{@{}>{\hspre}l<{\hspost}@{}}%
\column{23}{@{}>{\hspre}l<{\hspost}@{}}%
\column{E}{@{}>{\hspre}l<{\hspost}@{}}%
\>[B]{}\mathbf{data}\;\Conid{Opt'}\;\Varid{a}\mathrel{=}\Conid{Opt'}\;\{\mskip1.5mu \Varid{def}\mathbin{::}\Conid{Dp}\;\Conid{Bool},\Varid{val}\mathbin{::}\Varid{a}\mskip1.5mu\}{}\<[E]%
\\[\blanklineskip]%
\>[B]{}\mathbf{instance}\;\Conid{Syn}\;\Varid{a}\Rightarrow \Conid{Syn}\;(\Conid{Opt'}\;\Varid{a})\;\mathbf{where}{}\<[E]%
\\
\>[B]{}\hsindent{3}{}\<[3]%
\>[3]{}\mathbf{type}\;\Conid{Internal}\;(\Conid{Opt'}\;\Varid{a})\mathrel{=}(\Conid{Bool},\Conid{Internal}\;\Varid{a}){}\<[E]%
\\
\>[B]{}\hsindent{3}{}\<[3]%
\>[3]{}\Varid{toDp}\;(\Conid{Opt'}\;\Varid{b}\;\Varid{x}){}\<[20]%
\>[20]{}\mathrel{=}{}\<[23]%
\>[23]{}\Conid{Pair}\;\Varid{b}\;(\Varid{toDp}\;\Varid{x}){}\<[E]%
\\
\>[B]{}\hsindent{3}{}\<[3]%
\>[3]{}\Varid{fromDp}\;\Varid{p}{}\<[20]%
\>[20]{}\mathrel{=}{}\<[23]%
\>[23]{}\Conid{Opt'}\;(\Conid{Fst}\;\Varid{p})\;(\Varid{fromDp}\;(\Conid{Snd}\;\Varid{p})){}\<[E]%
\\[\blanklineskip]%
\>[B]{}\Varid{some'}{}\<[16]%
\>[16]{}\mathbin{::}{}\<[16E]%
\>[20]{}\Varid{a}\to \Conid{Opt'}\;\Varid{a}{}\<[E]%
\\
\>[B]{}\Varid{some'}\;\Varid{x}{}\<[16]%
\>[16]{}\mathrel{=}{}\<[16E]%
\>[20]{}\Conid{Opt'}\;\Varid{true}\;\Varid{x}{}\<[E]%
\\[\blanklineskip]%
\>[B]{}\Varid{none'}{}\<[16]%
\>[16]{}\mathbin{::}{}\<[16E]%
\>[20]{}\Conid{Undef}\;\Varid{a}\Rightarrow \Conid{Opt'}\;\Varid{a}{}\<[E]%
\\
\>[B]{}\Varid{none'}{}\<[16]%
\>[16]{}\mathrel{=}{}\<[16E]%
\>[20]{}\Conid{Opt'}\;\Varid{false}\;\Varid{undef}{}\<[E]%
\\[\blanklineskip]%
\>[B]{}\Varid{option'}{}\<[16]%
\>[16]{}\mathbin{::}{}\<[16E]%
\>[20]{}\Conid{Syn}\;\Varid{b}\Rightarrow \Varid{b}\to (\Varid{a}\to \Varid{b})\to \Conid{Opt'}\;\Varid{a}\to \Varid{b}{}\<[E]%
\\
\>[B]{}\Varid{option'}\;\Varid{d}\;\Varid{f}\;\Varid{o}{}\<[16]%
\>[16]{}\mathrel{=}{}\<[16E]%
\>[20]{}\Varid{def}\;\Varid{o}\mathbin{?}(\Varid{f}\;(\Varid{val}\;\Varid{o}),\Varid{d}){}\<[E]%
\ColumnHook
\end{hscode}\resethooks

The next obvious step is to define a suitable monad over the type \ensuremath{\Conid{Opt'}}.
The natural definitions to use are as follows:
\begin{hscode}\SaveRestoreHook
\column{B}{@{}>{\hspre}l<{\hspost}@{}}%
\column{11}{@{}>{\hspre}c<{\hspost}@{}}%
\column{11E}{@{}l@{}}%
\column{15}{@{}>{\hspre}l<{\hspost}@{}}%
\column{21}{@{}>{\hspre}l<{\hspost}@{}}%
\column{E}{@{}>{\hspre}l<{\hspost}@{}}%
\>[B]{}\Varid{return}{}\<[11]%
\>[11]{}\mathbin{::}{}\<[11E]%
\>[15]{}\Varid{a}\to \Conid{Opt'}\;\Varid{a}{}\<[E]%
\\
\>[B]{}\Varid{return}\;\Varid{x}{}\<[11]%
\>[11]{}\mathrel{=}{}\<[11E]%
\>[15]{}\Varid{some'}\;\Varid{x}{}\<[E]%
\\[\blanklineskip]%
\>[B]{}(\bind ){}\<[11]%
\>[11]{}\mathbin{::}{}\<[11E]%
\>[15]{}(\Conid{Undef}\;\Varid{b})\Rightarrow \Conid{Opt'}\;\Varid{a}\to (\Varid{a}\to \Conid{Opt'}\;\Varid{b})\to \Conid{Opt'}\;\Varid{b}{}\<[E]%
\\
\>[B]{}\Varid{o}\bind \Varid{g}{}\<[11]%
\>[11]{}\mathrel{=}{}\<[11E]%
\>[15]{}\Conid{Opt'}\;{}\<[21]%
\>[21]{}(\Varid{def}\;\Varid{o}\mathbin{?}(\Varid{def}\;(\Varid{g}\;(\Varid{val}\;\Varid{o})),\Varid{false}))\;{}\<[E]%
\\
\>[21]{}(\Varid{def}\;\Varid{o}\mathbin{?}(\Varid{val}\;(\Varid{g}\;(\Varid{val}\;\Varid{o})),\Varid{undef})){}\<[E]%
\ColumnHook
\end{hscode}\resethooks
However, this adds type constraint \ensuremath{\Conid{Undef}\;\Varid{b}}
to the type of \ensuremath{(\bind )}, which is not permitted.
The need to add such constraints often arises, and has
been dubbed the constrained-monad problem
\citep{hughes1999restricted,SvenningssonS13}.
We solve it with a trick due to \citet{PerssonAS11}.


We introduce a continuation-passing style (CPS) type, \ensuremath{\Conid{Opt}},
defined in terms of \ensuremath{\Conid{Opt'}}.  It is
straightforward to define \ensuremath{\Conid{Monad}} and \ensuremath{\Conid{Syn}} instances,
operations to lift the representation type to lift and lower
one type to the other, and to lift \ensuremath{\Varid{some}}, \ensuremath{\Varid{none}}, and
\ensuremath{\Varid{option}} to the CPS type.
The \ensuremath{\Varid{lift}} operation is closely related to the \ensuremath{(\bind )} operation
we could not define above; it is properly typed,
thanks to the type constraint on \ensuremath{\Varid{b}} in the definition of \ensuremath{\Conid{Opt}\;\Varid{a}}.

\begin{hscode}\SaveRestoreHook
\column{B}{@{}>{\hspre}l<{\hspost}@{}}%
\column{3}{@{}>{\hspre}l<{\hspost}@{}}%
\column{9}{@{}>{\hspre}l<{\hspost}@{}}%
\column{11}{@{}>{\hspre}l<{\hspost}@{}}%
\column{14}{@{}>{\hspre}l<{\hspost}@{}}%
\column{15}{@{}>{\hspre}c<{\hspost}@{}}%
\column{15E}{@{}l@{}}%
\column{18}{@{}>{\hspre}l<{\hspost}@{}}%
\column{26}{@{}>{\hspre}l<{\hspost}@{}}%
\column{E}{@{}>{\hspre}l<{\hspost}@{}}%
\>[B]{}\mathbf{newtype}\;\Conid{Opt}\;\Varid{a}\mathrel{=}{}\<[E]%
\\
\>[B]{}\hsindent{3}{}\<[3]%
\>[3]{}\Conid{O}\;\{\mskip1.5mu \Varid{unO}\mathbin{::}\forall \Varid{b}\hsforall \hsdot{\circ }{.}\Conid{Undef}\;\Varid{b}\Rightarrow ((\Varid{a}\to \Conid{Opt'}\;\Varid{b})\to \Conid{Opt'}\;\Varid{b})\mskip1.5mu\}{}\<[E]%
\\[\blanklineskip]%
\>[B]{}\mathbf{instance}\;\Conid{Monad}\;\Conid{Opt}\;\mathbf{where}{}\<[E]%
\\
\>[B]{}\hsindent{3}{}\<[3]%
\>[3]{}\Varid{return}\;\Varid{x}{}\<[15]%
\>[15]{}\mathrel{=}{}\<[15E]%
\>[18]{}\Conid{O}\;(\lambda \Varid{g}\to \Varid{g}\;\Varid{x}){}\<[E]%
\\
\>[B]{}\hsindent{3}{}\<[3]%
\>[3]{}\Varid{m}\bind \Varid{k}{}\<[15]%
\>[15]{}\mathrel{=}{}\<[15E]%
\>[18]{}\Conid{O}\;(\lambda \Varid{g}\to \Varid{unO}\;\Varid{m}\;(\lambda \Varid{x}\to \Varid{unO}\;(\Varid{k}\;\Varid{x})\;\Varid{g})){}\<[E]%
\\[\blanklineskip]%
\>[B]{}\mathbf{instance}\;\Conid{Undef}\;\Varid{a}\Rightarrow \Conid{Syn}\;(\Conid{Opt}\;\Varid{a})\;\mathbf{where}{}\<[E]%
\\
\>[B]{}\hsindent{3}{}\<[3]%
\>[3]{}\mathbf{type}\;\Conid{Internal}\;(\Conid{Opt}\;\Varid{a})\mathrel{=}(\Conid{Bool},\Conid{Internal}\;\Varid{a}){}\<[E]%
\\
\>[B]{}\hsindent{3}{}\<[3]%
\>[3]{}\Varid{fromDp}{}\<[11]%
\>[11]{}\mathrel{=}{}\<[14]%
\>[14]{}\Varid{lift}\hsdot{\circ }{.}\Varid{fromDp}{}\<[E]%
\\
\>[B]{}\hsindent{3}{}\<[3]%
\>[3]{}\Varid{toDp}{}\<[11]%
\>[11]{}\mathrel{=}{}\<[14]%
\>[14]{}\Varid{toDp}\hsdot{\circ }{.}\Varid{lower}{}\<[E]%
\\[\blanklineskip]%
\>[B]{}\Varid{lift}\mathbin{::}\Conid{Opt'}\;\Varid{a}\to \Conid{Opt}\;\Varid{a}{}\<[E]%
\\
\>[B]{}\Varid{lift}\;\Varid{o}\mathrel{=}{}\<[11]%
\>[11]{}\Conid{O}\;(\lambda \Varid{g}\to \Conid{Opt'}\;{}\<[26]%
\>[26]{}(\Varid{def}\;\Varid{o}\mathbin{?}(\Varid{def}\;(\Varid{g}\;(\Varid{val}\;\Varid{o})),\Varid{false}))\;{}\<[E]%
\\
\>[26]{}(\Varid{def}\;\Varid{o}\mathbin{?}(\Varid{val}\;(\Varid{g}\;(\Varid{val}\;\Varid{o})),\Varid{undef}))){}\<[E]%
\\[\blanklineskip]%
\>[B]{}\Varid{lower}\mathbin{::}\Conid{Undef}\;\Varid{a}\Rightarrow \Conid{Opt}\;\Varid{a}\to \Conid{Opt'}\;\Varid{a}{}\<[E]%
\\
\>[B]{}\Varid{lower}\;\Varid{m}\mathrel{=}\Varid{unO}\;\Varid{m}\;\Varid{some'}{}\<[E]%
\\[\blanklineskip]%
\>[B]{}\Varid{none}\mathbin{::}\Conid{Undef}\;\Varid{a}\Rightarrow \Conid{Opt}\;\Varid{a}{}\<[E]%
\\
\>[B]{}\Varid{none}\mathrel{=}{}\<[9]%
\>[9]{}\Varid{lift}\;\Varid{none'}{}\<[E]%
\\[\blanklineskip]%
\>[B]{}\Varid{some}\mathbin{::}\Varid{a}\to \Conid{Opt}\;\Varid{a}{}\<[E]%
\\
\>[B]{}\Varid{some}\;\Varid{a}\mathrel{=}\Varid{lift}\;(\Varid{some'}\;\Varid{a}){}\<[E]%
\\[\blanklineskip]%
\>[B]{}\Varid{option}\mathbin{::}(\Conid{Undef}\;\Varid{a},\Conid{Syn}\;\Varid{b})\Rightarrow \Varid{b}\to (\Varid{a}\to \Varid{b})\to \Conid{Opt}\;\Varid{a}\to \Varid{b}{}\<[E]%
\\
\>[B]{}\Varid{option}\;\Varid{d}\;\Varid{f}\;\Varid{o}\mathrel{=}\Varid{option'}\;\Varid{d}\;\Varid{f}\;(\Varid{lower}\;\Varid{o}){}\<[E]%
\ColumnHook
\end{hscode}\resethooks
These definitions support the EDSL code presented
in Section~\ref{subsec:e-maybe}.

\subsection{Embedding vector}

Recall that values of type \ensuremath{\Conid{Array}} are created by construct \ensuremath{\Conid{MkArr}},
while \ensuremath{\Conid{LnArr}} extracts the length and \ensuremath{\Conid{IxArr}} fetches the element at
the given index.  Corresponding to the deep embedding \ensuremath{\Conid{Array}} is a
shallow embedding \ensuremath{\Conid{Vec}}.
\begin{hscode}\SaveRestoreHook
\column{B}{@{}>{\hspre}l<{\hspost}@{}}%
\column{3}{@{}>{\hspre}l<{\hspost}@{}}%
\column{21}{@{}>{\hspre}c<{\hspost}@{}}%
\column{21E}{@{}l@{}}%
\column{24}{@{}>{\hspre}l<{\hspost}@{}}%
\column{29}{@{}>{\hspre}l<{\hspost}@{}}%
\column{E}{@{}>{\hspre}l<{\hspost}@{}}%
\>[B]{}\mathbf{data}\;\Conid{Vec}\;\Varid{a}\mathrel{=}\Conid{Vec}\;(\Conid{Dp}\;\Conid{Int})\;(\Conid{Dp}\;\Conid{Int}\to \Varid{a}){}\<[E]%
\\[\blanklineskip]%
\>[B]{}\mathbf{instance}\;\Conid{Syn}\;\Varid{a}\Rightarrow \Conid{Syn}\;(\Conid{Vec}\;\Varid{a})\;\mathbf{where}{}\<[E]%
\\
\>[B]{}\hsindent{3}{}\<[3]%
\>[3]{}\mathbf{type}\;\Conid{Internal}\;(\Conid{Vec}\;\Varid{a})\mathrel{=}\Conid{Array}\;\Conid{Int}\;(\Conid{Internal}\;\Varid{a}){}\<[E]%
\\
\>[B]{}\hsindent{3}{}\<[3]%
\>[3]{}\Varid{toDp}\;(\Conid{Vec}\;\Varid{n}\;\Varid{g}){}\<[21]%
\>[21]{}\mathrel{=}{}\<[21E]%
\>[24]{}\Conid{MkArr}\;\Varid{n}\;(\Varid{toDp}\hsdot{\circ }{.}\Varid{g}){}\<[E]%
\\
\>[B]{}\hsindent{3}{}\<[3]%
\>[3]{}\Varid{fromDp}\;\Varid{a}{}\<[21]%
\>[21]{}\mathrel{=}{}\<[21E]%
\>[24]{}\Conid{Vec}\;{}\<[29]%
\>[29]{}(\Conid{LnArr}\;\Varid{a})\;(\Varid{fromDp}\hsdot{\circ }{.}\Conid{IxArr}\;\Varid{a}){}\<[E]%
\\[\blanklineskip]%
\>[B]{}\mathbf{instance}\;\Conid{Functor}\;\Conid{Vec}\;\mathbf{where}{}\<[E]%
\\
\>[B]{}\hsindent{3}{}\<[3]%
\>[3]{}\Varid{fmap}\;\Varid{f}\;(\Conid{Vec}\;\Varid{n}\;\Varid{g}){}\<[21]%
\>[21]{}\mathrel{=}{}\<[21E]%
\>[24]{}\Conid{Vec}\;\Varid{n}\;(\Varid{f}\hsdot{\circ }{.}\Varid{g}){}\<[E]%
\ColumnHook
\end{hscode}\resethooks
Constructor \ensuremath{\Conid{Vec}} resembles \ensuremath{\Conid{Arr}}, but the former
constructs a high-level representation of the array and the latter an
actual array.
It is straightforward to make \ensuremath{\Conid{Vec}} an instance of \ensuremath{\Conid{Functor}}.

It is easy to define operations on vectors,
including combining corresponding elements of two vectors,
summing the elements of a vector, dot product of two vectors,
and norm of a vector.
\begin{hscode}\SaveRestoreHook
\column{B}{@{}>{\hspre}l<{\hspost}@{}}%
\column{9}{@{}>{\hspre}c<{\hspost}@{}}%
\column{9E}{@{}l@{}}%
\column{13}{@{}>{\hspre}l<{\hspost}@{}}%
\column{15}{@{}>{\hspre}l<{\hspost}@{}}%
\column{17}{@{}>{\hspre}l<{\hspost}@{}}%
\column{E}{@{}>{\hspre}l<{\hspost}@{}}%
\>[B]{}\Varid{zipVec}{}\<[9]%
\>[9]{}\mathbin{::}{}\<[9E]%
\>[13]{}(\Conid{Syn}\;\Varid{a},\Conid{Syn}\;\Varid{b})\Rightarrow {}\<[E]%
\\
\>[13]{}\hsindent{2}{}\<[15]%
\>[15]{}(\Varid{a}\to \Varid{b}\to \Varid{c})\to \Conid{Vec}\;\Varid{a}\to \Conid{Vec}\;\Varid{b}\to \Conid{Vec}\;\Varid{c}{}\<[E]%
\\
\>[B]{}\Varid{zipVec}\;\Varid{f}\;(\Conid{Vec}\;\Varid{m}\;\Varid{g})\;(\Conid{Vec}\;\Varid{n}\;\Varid{h}){}\<[E]%
\\
\>[B]{}\hsindent{9}{}\<[9]%
\>[9]{}\mathrel{=}{}\<[9E]%
\>[13]{}\Conid{Vec}\;(\Varid{m}\mathbin{`\Varid{minim}`}\Varid{n})\;(\lambda \Varid{i}\to \Varid{f}\;(\Varid{g}\;\Varid{i})\;(\Varid{h}\;\Varid{i})){}\<[E]%
\\[\blanklineskip]%
\>[B]{}\Varid{sumVec}{}\<[9]%
\>[9]{}\mathbin{::}{}\<[9E]%
\>[13]{}(\Conid{Syn}\;\Varid{a},\Conid{Num}\;\Varid{a})\Rightarrow \Conid{Vec}\;\Varid{a}\to \Varid{a}{}\<[E]%
\\
\>[B]{}\Varid{sumVec}\;(\Conid{Vec}\;\Varid{n}\;\Varid{g}){}\<[E]%
\\
\>[B]{}\hsindent{9}{}\<[9]%
\>[9]{}\mathrel{=}{}\<[9E]%
\>[13]{}\Varid{for}\;\Varid{n}\;\mathrm{0}\;(\lambda \Varid{i}\;\Varid{x}\to \Varid{x}\mathbin{+}\Varid{g}\;\Varid{i}){}\<[E]%
\\[\blanklineskip]%
\>[B]{}\Varid{dotVec}{}\<[13]%
\>[13]{}\mathbin{::}{}\<[17]%
\>[17]{}(\Conid{Syn}\;\Varid{a},\Conid{Num}\;\Varid{a})\Rightarrow \Conid{Vec}\;\Varid{a}\to \Conid{Vec}\;\Varid{a}\to \Varid{a}{}\<[E]%
\\
\>[B]{}\Varid{dotVec}\;\Varid{u}\;\Varid{v}{}\<[13]%
\>[13]{}\mathrel{=}{}\<[17]%
\>[17]{}\Varid{sumVec}\;(\Varid{zipVec}\;(\times)\;\Varid{u}\;\Varid{v}){}\<[E]%
\\[\blanklineskip]%
\>[B]{}\Varid{normVec}{}\<[13]%
\>[13]{}\mathbin{::}{}\<[17]%
\>[17]{}\Conid{Vec}\;(\Conid{Dp}\;\Conid{Float})\to \Conid{Dp}\;\Conid{Float}{}\<[E]%
\\
\>[B]{}\Varid{normVec}\;\Varid{v}{}\<[13]%
\>[13]{}\mathrel{=}{}\<[17]%
\>[17]{}\Varid{sqrt}\;(\Varid{dotVec}\;\Varid{v}\;\Varid{v}){}\<[E]%
\ColumnHook
\end{hscode}\resethooks
Invoking \ensuremath{\Varid{edsl}} on
\[
\ensuremath{\Varid{normVec}\hsdot{\circ }{.}\Varid{toVec}}
\]
generates C code to normalise a vector. If we used a top-level function
of type \ensuremath{(\Conid{Syn}\;\Varid{a},\Conid{Syn}\;\Varid{b})\Rightarrow (\Varid{a}\to \Varid{b})\to \Conid{C}}, then it would insert the
\ensuremath{\Varid{toVec}} coercion automatically.

This style of definition again provides fusion. For instance:
\[
\begin{array}{cl}
  &  \ensuremath{\Varid{dotVec}\;(\Conid{Vec}\;\Varid{m}\;\Varid{g})\;(\Conid{Vec}\;\Varid{n}\;\Varid{h})}  \\
= &  \ensuremath{\Varid{sumVec}\;(\Varid{zipVec}\;(\times)\;(\Conid{Vec}\;\Varid{m}\;\Varid{g})\;(\Conid{Vec}\;\Varid{n}\;\Varid{h})}  \\
= &  \ensuremath{\Varid{sumVec}\;(\Conid{Vec}\;(\Varid{m}\mathbin{`\Varid{minim}`}\Varid{n})\;(\lambda \Varid{i}\to \Varid{g}\;\Varid{i}\times\Varid{h}\;\Varid{i})}  \\
= &  \ensuremath{\Varid{for}\;(\Varid{m}\mathbin{`\Varid{minim}`}\Varid{n})\;(\lambda \Varid{i}\;\Varid{x}\to \Varid{x}\mathbin{+}\Varid{g}\;\Varid{i}\times\Varid{h}\;\Varid{i})}
\end{array}
\]
Indeed, we can see that by construction that whenever we combine two
primitives the intermediate vector is always eliminated.

The type class \ensuremath{\Conid{Syn}} enables conversion between types \ensuremath{\Conid{Arr}} and \ensuremath{\Conid{Vec}}.
Hence for EDSL, unlike QDSL, explicit calls \ensuremath{\Varid{toVec}} and \ensuremath{\Varid{fromVec}} are
not required.  Invoking \ensuremath{\Varid{edsl}\;\Varid{normVec}} produces the same C code as in
Section~\ref{subsec:arrays}.

As with QDSL, there are some situations where fusion is not beneficial.
We may materialise a vector as an array with the following function.
\begin{hscode}\SaveRestoreHook
\column{B}{@{}>{\hspre}l<{\hspost}@{}}%
\column{E}{@{}>{\hspre}l<{\hspost}@{}}%
\>[B]{}\Varid{memorise}\mathbin{::}\Conid{Syn}\;\Varid{a}\Rightarrow \Conid{Vec}\;\Varid{a}\to \Conid{Vec}\;\Varid{a}{}\<[E]%
\\
\>[B]{}\Varid{memorise}\mathrel{=}\Varid{fromDp}\hsdot{\circ }{.}\Conid{Save}\hsdot{\circ }{.}\Varid{toDp}{}\<[E]%
\ColumnHook
\end{hscode}\resethooks
Here we interpose \ensuremath{\Conid{Save}} to forestall the fusion that would otherwise occur.
For example, if
\begin{hscode}\SaveRestoreHook
\column{B}{@{}>{\hspre}l<{\hspost}@{}}%
\column{18}{@{}>{\hspre}l<{\hspost}@{}}%
\column{E}{@{}>{\hspre}l<{\hspost}@{}}%
\>[B]{}\Varid{blur}\mathbin{::}\Conid{Syn}\;\Varid{a}\Rightarrow \Conid{Vec}\;\Varid{a}\to \Conid{Vec}\;\Varid{a}{}\<[E]%
\\
\>[B]{}\Varid{blur}\;\Varid{v}\mathrel{=}\Varid{zipVec}\;{}\<[18]%
\>[18]{}(\lambda \Varid{x}\;\Varid{y}\to \Varid{sqrt}\;(\Varid{x}\times\Varid{y}))\;{}\<[E]%
\\
\>[18]{}(\Varid{appVec}\;\Varid{a}\;(\Varid{uniVec}\;\mathrm{0}))\;{}\<[E]%
\\
\>[18]{}(\Varid{appVec}\;(\Varid{uniVec}\;\mathrm{0})\;\Varid{a}){}\<[E]%
\ColumnHook
\end{hscode}\resethooks
computes the geometric mean of adjacent elements of a vector, then one may choose to
compute either
\begin{center}
\ensuremath{\Varid{blur}\hsdot{\circ }{.}\Varid{blur}} ~~~or~~~ \ensuremath{\Varid{blur}\hsdot{\circ }{.}\Varid{memorise}\hsdot{\circ }{.}\Varid{blur}}
\end{center}
with different trade-offs between recomputation and memory use.

QDSL forces all conversions to be written out, while EDSL silently
converts between representations; following the pattern that QDSL is
more explicit, while EDSL is more compact.  For QDSL it is the
subformula property which guarantees that all intermediate uses of
\ensuremath{\Conid{Vec}} are eliminated, while for EDSL this is established by
operational reasoning on the behaviour of the type \ensuremath{\Conid{Vec}}.

\section{Related work}
\label{sec:related}

DSLs have a long and rich history \citep{Bentley-1986}.
An early use of quotation in programming is Lisp \citep{McCarthy-1960},
and perhaps the first application of quotation to domain-specific
languages is Lisp macros \citep{Hart-1963}.

This paper uses Haskell, which has been widely used for EDSLs
\citep{hudak1997domain}.
We constrast QDSL with an EDSL technique
that combines deep and shallow embedding, as described by
\citet{svenningsson:combining}, and as used in several Haskell EDSLs
including Feldspar \citep{FELDSPAR}, Obsidian
\citep{svensson2011obsidian}, Nikola \citep{NIKOLA}, Hydra
\citep{giorgidze2011embedding}, and Meta-Repa \citep{ankner2013edsl}.


\citet{odonnell:sharing} identified loss of sharing in the context of
embedded circuit descriptions.  \citet{claessen1999observable} extended
Haskell to support observable sharing.  \citet{gill2009type} proposes
library features that support sharing without need to extend the
language.

A proposition-as-types principle for quotation as a modal logic was
proposed by \citet{Davies-Pfenning-2001}.  As
they note, their technique has close connections to two-level
languages \citep{Nielson-2005} and partial evaluation
\citep{jones1993partial}.

Other approaches to DSL that make use of quotation include
C\# and F\# versions of LINQ
\citep{csharplinq,fsharplinq} and Scala Lightweight
Modular Staging (LMS) \citep{scalalms}.
Scala LMS exploits techniques found in both QDSL
(quotation and normalisation)
and EDSL (combining shallow and deep embeddings),
see \citet{rompf2013scala}, and exploits reuse to
allow multiple DSLs to share infrastructure
see \citet{sujeeth2013composition}.

The underlying idea for QDSLs was established
for F\# LINQ by \citet{cheney:linq}, and extended
to nested results by \citet{CheneyLW14}.
Related work combines language-integrated query
with effect types \citep{Cooper09,LindleyC12}.
\citet{CheneyLRW14} compare approaches
based on quotation and effects.

\section{Conclusion}
\label{sec:conclusion}

\begin{quote}
A good idea can be much better than a new one. \\
\flushr -- Gerard Berry
\end{quote}

We have compared QDSLs and EDSLs, arguing that QDSLs offer competing
expressiveness and efficiency.

The subformula property may have applications in DSLs other that
QDSLs. For instance, after Section~\ref{subsec:opt} of this paper was
drafted, it occurred to us that a different approach
would be to extend type \ensuremath{\Conid{Dp}} with constructs for type \ensuremath{\Conid{Maybe}}.
So long as type \ensuremath{\Conid{Maybe}} does not appear in the input or output of the
program, a normaliser that ensures the subformula property could
guarantee that C code for such constructs need never be generated.

Rather than building a special-purpose tool for each QDSL, it should
be possible to design a single tool for each host language.  Our next
step is to design a QDSL library for Haskell that restores the type
information for quasi-quotations currencly discarded by GHC and uses
this to support type classes and overloading in full, and to supply a
more general normaliser.  Such a tool would subsume the
special-purpose translator from \ensuremath{\Conid{Qt}} to \ensuremath{\Conid{Dp}} described at the
beginning of Section~\ref{sec:implementation}, and lift most of its
restrictions.

\begin{quote}
  These forty years now I've been
  speaking in prose without knowing it!
  \flushr --- Moliere
\end{quote}

Like Moli\`{e}re's Monsieur Jourdain, many of us have used QDSLs for
years, if not by that name. DSL via quotation lies at the heart of Lisp
macros, Microsoft LINQ, and Scala LMS, to name but three.
By naming the concept and drawing attention to the benefits of
normalisation and the subformula propety, we hope to help the concept to
prosper for years to come.

\paragraph*{Acknowledgement}
Najd is supported by a Google Europe Fellowship in Programming
Technology.
Svenningsson is a SICSA Visiting Fellow and is funded by a HiPEAC
collaboration grant, and by the Swedish Foundation for Strategic
Research under grant RawFP.
Lindley and Wadler are funded by EPSRC Grant EP/K034413/1.

\bibliographystyle{plainnat}
\bibliography{paper}

\end{document}